\documentclass{article}%
\usepackage{amsfonts}
\usepackage{amsmath}%
\usepackage{amssymb}%
\usepackage{graphicx}
\providecommand{\U}[1]{\protect\rule{.1in}{.1in}}

\begin{document}

\title{Direct derivation of the comma 3-Vertex in the full string basis}
\author{A Abdurrahman, I Abdurrahman\thanks{Departmen of Physics, University of
Washington, 5720 South Ellis Avenue, Seattle, WA 60637, USA; ia4021@uw.edu}
and M Gassem\thanks{The Division of Mathematics and Science, South Texas
College, 3201 W. Pecan, McAllen, Texas 78501; mgassem@southtexascollege.edu}\\Department of Physics\\Shippensburg University of Pennsylvania\\1871 Old Main Drive\\Shippensburg, PA 17257\\USA}
\maketitle

\begin{abstract}
The interacting comma 3-vertex for the bosonic open string in the full string
basis is derived using the half string overlap relations directly. Thus
avoiding the coherent states technique employed in earlier derivations. The
resulting form of the interacting 3-vertex turns out to be precisely the
desired expression obtained in terms of the full string oscillator modes. This
derivation establishes that the comma 3-vertex and Witten's 3-vertex are
identical and therefore are interchangeable.

\end{abstract}

\section{Introduction}

Here we are going to give a brief derivation of the transformation matrices
between the half string coordinates and the full string coordinates needed for
the construction of the half string interacting vertex in terms of the
oscillator representation of the full string. For this we shall follow closely
the discussion of reference \cite{1,2,3,4,5}. To make this more concrete we
recall the standard mode expansion for the open bosonic string coordinate%
\begin{equation}
x^{\mu}(\sigma)=x_{0}^{\mu}+\sqrt{2}\sum_{n=1}^{\infty}x_{n}^{\mu}\cos
(n\sigma),\text{ \ \ \ }\sigma\in\left[  0,\pi\right] \label{eqnhscoordn}%
\end{equation}
where $\mu=1,2,....,27$ and $x^{27}(\sigma)$ correspond to the ghost part
$\phi\left(  \sigma\right)  $. The half string coordinates $x^{L,\mu}(\sigma)$
and $x^{R,\mu}(\sigma)$ for the left and right halves of the string are
defined in the usual way%
\begin{align}
x^{L,\mu}(\sigma)  & =x^{\mu}(\sigma)-x^{\mu}(\frac{\pi}{2})\text{,
\ \ \ }\sigma\in\left[  0,\frac{\pi}{2}\right] \nonumber\\
x^{R,\mu}(\sigma)  & =x^{\mu}(\pi-\sigma)-x^{\mu}(\frac{\pi}{2})\text{,
\ \ \ }\sigma\in\left[  0,\frac{\pi}{2}\right] \label{eqdefofhscoorn}%
\end{align}
where both $x^{L,\mu}(\sigma)$ and $x^{R,\mu}(\sigma)$ satisfy the usual
Neumann boundary conditions at $\sigma=0$ and a Dirichlet boundary conditions
$\sigma=\pi/2$. Thus they have expansions of the form%
\begin{align}
x^{L,\mu}(\sigma)  & =\sqrt{2}\sum_{n=1}^{\infty}x_{n}^{L\mu}\cos
(n\sigma)\text{,}\nonumber\\
x^{R,\mu}(\sigma)  & =\sqrt{2}\sum_{n=1}^{\infty}x_{n}^{R\mu}\cos
(n\sigma)\label{eqndefofhscoor}%
\end{align}
Comparing equation (\ref{eqnhscoordn}) and equation (\ref{eqndefofhscoor}) we
obtain an expression for the half string modes in terms of the full string
modes%
\begin{align}
x_{n}^{L,\mu}  & =x_{2n-1}^{\mu}+\sum_{m=1}^{\infty}\sqrt{\frac{2m}{2n-1}%
}\left[  M_{mn}^{1}+M_{mn}^{2}\right]  x_{2m}^{\mu}\text{,}\nonumber\\
x_{n}^{R,\mu}  & =-x_{2n-1}^{\mu}+\sum_{m=1}^{\infty}\sqrt{\frac{2m}{2n-1}%
}\left[  M_{mn}^{1}+M_{mn}^{2}\right]  x_{2m}^{\mu}\label{eqntransfbtwfh}%
\end{align}
where the change of representation matrices are given by%
\begin{equation}
M_{m\text{ }n\text{ }}^{1}=\frac{2}{\pi}\sqrt{\frac{2m}{2n-1}}\frac{\left(
-1\right)  ^{m+n}}{2m-\left(  2n-1\right)  },\text{ \ \ }m,n=1,2,3,...
\end{equation}%
\begin{equation}
M_{m\text{ }n\text{ }}^{2}=\frac{2}{\pi}\sqrt{\frac{2m}{2n-1}}\frac{\left(
-1\right)  ^{m+n}}{2m+\left(  2n-1\right)  }\text{ \ \ }m,n=1,2,3,...
\end{equation}
Since the transformation in (\ref{eqntransfbtwfh}) is non singular, one may
invert the relation in (\ref{eqntransfbtwfh}). Inverting (\ref{eqntransfbtwfh}%
) we find%
\begin{align}
x_{2n-1}^{\mu}  & =\frac{1}{2}\left(  x_{n}^{L,\mu}-x_{n}^{R,\mu}\right)
,\text{\ \ \ }\nonumber\\
x_{2n}^{\mu}  & =\frac{1}{2}\sum_{m=1}^{\infty}\sqrt{\frac{2m-1}{2n}}\left[
M_{mn}^{1}-M_{mn}^{2}\right]  \left(  x_{m}^{L,\mu}+x_{m}^{R,\mu}\right)
\label{eqntranshbtwf}%
\end{align}
where $n=1,2,3,...$.

In the decomposition of the string into right and left pieces in
(\ref{eqdefofhscoorn}), we singled out the midpoint coordinate. Consequently
the relationship between $x_{n}^{\mu}$ and $\left(  x_{n}^{L,\mu},x_{n}%
^{R,\mu}\right)  $ does not involve the zero mode $x_{0}^{\mu}$ \ of $x^{\mu
}(\sigma)$. At $\sigma=\pi/2$, we have%
\begin{equation}
x_{M}^{\mu}\equiv x^{\mu}\left(  \frac{\pi}{2}\right)  =x_{0}^{\mu}+\sqrt
{2}\sum_{n=1}^{\infty}x_{2n}^{\mu}\label{eqnmidtoze}%
\end{equation}
and so the center of mass $x_{0}^{\mu}$ may be related to the half string
coordinates and the midpoint coordinate%
\begin{equation}
x_{0}^{\mu}=x_{M}^{\mu}-\frac{\sqrt{2}}{\pi}\sum_{n=1}^{\infty}\frac{\left(
-1\right)  ^{n}}{2n-1}\left(  x_{n}^{L,\mu}+x_{n}^{R,\mu}\right)
\label{eqnzetomid}%
\end{equation}
Equations (\ref{eqnmidtoze}) and (\ref{eqnzetomid}) with equations
(\ref{eqntransfbtwfh}) and (\ref{eqntranshbtwf}) complete the equivalence
between $x_{n}^{\mu}$, $n=0,1,2...$, and $\left(  x_{n}^{L,\mu},x_{n}^{R,\mu
},x_{M}^{\mu}\right)  $, $n=1,2,3,...$.

For later use we also need the relationships between$\left\{  p_{n}^{L,\mu
},p_{n}^{R,\mu},p_{M}^{\mu}\right\}  _{n=1}^{\infty}$, the half string
conjugate momenta and $\left\{  p_{n}^{\mu}\right\}  _{n=0}^{\infty}$, the
full string conjugate momenta . Using Dirac quantization procedure%
\begin{equation}
\left[  x_{n}^{r},p_{m}^{s}\right]  =i\delta^{rs}\delta_{nm}\text{,}%
\end{equation}
we find (thereafter; the space-time index $\mu$ is suppressed),%
\begin{align}
p_{n}^{L}  & =\frac{1}{2}p_{2n-1}+\sum_{m=1}^{\infty}\sqrt{\frac{2n-1}{2m}%
}\left[  M_{mn}^{1}-M_{mn}^{2}\right]  p_{2m}\nonumber\\
& -\frac{\sqrt{2}}{\pi}\frac{\left(  -1\right)  ^{n}}{2n-1}p_{0}\text{,}\\
p_{n}^{R}  & =-\frac{1}{2}p_{2n-1}+\sum_{m=1}^{\infty}\sqrt{\frac{2n-1}{2m}%
}\left[  M_{mn}^{1}-M_{mn}^{2}\right]  p_{2m}\nonumber\\
& -\frac{\sqrt{2}}{\pi}\frac{\left(  -1\right)  ^{n}}{2n-1}p_{0}%
\end{align}
and%
\begin{equation}
p_{_{M}}=p_{0}%
\end{equation}
To obtain the full string conjugate momenta in terms of the half string
conjugate momenta, we need to invert the above relations; skipping the
technical details we find%
\begin{align}
p_{2n-1}  & =p_{n}^{L}-p_{n}^{R},\text{\ \ \ }\nonumber\\
p_{2n}  & =\sum_{m=1}^{\infty}\sqrt{\frac{2n}{2m-1}}\left[  M_{nm}^{1}%
+M_{nm}^{2}\right]  \left(  p_{m}^{L}+p_{m}^{R}\right)  +\sqrt{2}\left(
-1\right)  ^{n}p_{M}%
\end{align}
We notice that the existence of the one-to-one correspondence between the half
string and the full string degrees of freedom guarantees the existence of the
identification%
\begin{equation}
\overline{H}=\overline{H_{M}\otimes H_{L}\otimes H_{R}}%
\end{equation}
where $\overline{H}$ stands for the completion of the full string Hilbert
space and $H_{L}$, $H_{R}$, $H_{M}$ in the tensor product stand for the two
half-string Hilbert spaces and the Hilbert space of functions of the
mid-point, respectively.

\section{\textbf{The half-string overlaps}}

The half string three interaction vertex of the open bosonic string
($V_{x}^{HS}$) have been constructed in the half-string oscillator
representation \cite{2,3}. Here we are interested in constructing the comma
three interaction vertex in terms of the oscillator representation of the full
string. Here we shall only consider the coordinate piece of the comma three
interaction vertex. The ghost part of the vertex ($V_{\phi}^{HS}$) in the
bosonic representation is identical to the coordinate piece apart from the
ghost mid-point insertions $3i\phi\left(  \pi/2\right)  /2$ \ required for
ghost number conservation at the mid-point. To simplify the calculation we
introduce a new set of coordinates and momenta based on a $Z_{3}$ Fourier
transform\footnote{This technique was first used by D. Gross and A. Jevicki in
1986.}%
\begin{equation}
\left(
\begin{array}
[c]{c}%
Q^{r}\left(  \sigma\right) \\
\overline{Q}^{r}\left(  \sigma\right) \\
Q^{3,r}\left(  \sigma\right)
\end{array}
\right)  =\frac{1}{\sqrt{3}}\left(
\begin{array}
[c]{ccc}%
e & \bar{e} & 1\\
\bar{e} & e & 1\\
1 & 1 & 1
\end{array}
\right)  \left(
\begin{array}
[c]{c}%
\chi^{1,r}\left(  \sigma\right) \\
\chi^{2,r}\left(  \sigma\right) \\
\chi^{3,r}\left(  \sigma\right)
\end{array}
\right) \label{Z3transform}%
\end{equation}
where $e=\exp\left(  2\pi i/3\right)  $ and $r$ refers to the left ($L$) and
right ($R$) parts of the string. The superscripts $1$, $2$ and $3$ refers to
string $1$, string $2$ and string $3$, respectively. Similarly one obtains a
new set for the conjugate momenta  $\wp^{r}\left(  \sigma\right)  $,
$\overline{\wp}^{r}\left(  \sigma\right)  $ and  $\wp^{3,r}\left(
\sigma\right)  $ as well as a new set for the creation-annihilation operators
$\left(  B_{j}^{r},B_{j}^{r\dag}\right)  $. In the $Z_{3}$ Fourier space the
degrees of freedom in the $\delta$ function overlaps equations decouple which
result in a considerable reduction of the amount of algebra involved in such
calculations as we shall see shortly. Notice that in the $Z_{3}$ Fourier space
the commutation relations are
\begin{align}
\left[  Q^{r}\left(  \sigma\right)  ,\overline{\wp }^{s}\left(
\sigma^{\prime}\right)  \right]   & =i\delta^{rs}\delta\left(  \sigma
-\sigma^{\prime}\right) \\
\left[  \overline{Q}^{r}\left(  \sigma\right)  ,\wp ^{s}\left(
\sigma^{\prime}\right)  \right]   & =i\delta^{rs}\delta\left(  \sigma
-\sigma^{\prime}\right) \\
\left[  Q^{3,r}\left(  \sigma\right)  ,\wp ^{3,s}\left(  \sigma
^{\prime}\right)  \right]   & =i\delta^{rs}\delta\left(  \sigma-\sigma
^{\prime}\right)
\end{align}
Since $\left[  Q^{r}\left(  \sigma\right)  ,\wp ^{s}\left(
\sigma^{\prime}\right)  \right]  \neq i\delta^{rs}\delta\left(  \sigma
-\sigma^{\prime}\right)  $, then $Q^{r}\left(  \sigma\right)  $ and
 $\wp^{r}\left(  \sigma\right)  $ are no longer canonical variables. The
canonical variables in this case are $Q^{r}\left(  \sigma\right)  $ and
$\overline{\wp }^{r}\left(  \sigma\right)  $. Thus the $Z_{3}$ Fourier
transform does not conserve the original commutation relations. The variables
$Q^{3,r}\left(  \sigma\right)  $ and  $\wp^{3,s}\left(  \sigma\right)  $ are
still canonical however. This is a small price to pay for decoupling string
three in the $Z_{3}$ Fourier space from the other two strings as we shall see
in the construction of the comma \ three interaction vertex. Recall that the
overlap equations for the comma three interacting vertex are given by%
\begin{align}
\chi^{j,r}\left(  \sigma\right)   & =\chi^{j-1,r-1}\left(  \sigma\right)
,\text{ \ \ \ \ \ \ }0\leq\sigma\leq\pi/2\\
x_{M}^{1}  & =x_{M}^{2}=x_{M}^{3}\label{eqnmidpconx}%
\end{align}
for the coordinates (where the mid-point coordinate $x_{M}\equiv x\left(
\pi/2\right)  $ and the identifications $j-1=0\equiv3$ and $r-1=0\equiv R$ are
understood). The comma coordinates are defined in the usual way \cite{1}%
\begin{align}
\chi^{j,L}\left(  \sigma\right)   & =x\left(  \sigma\right)  -x\left(
\frac{\pi}{2}\right)  ,\text{\ \ \ \ \ \ \ }0\leq\sigma\leq\pi/2\\
\chi^{j,R}\left(  \sigma\right)   & =x\left(  \pi-\sigma\right)  -x\left(
\frac{\pi}{2}\right)  ,\text{\ \ \ \ \ \ \ }0\leq\sigma\leq\pi/2
\end{align}
The overlaps for the canonical momenta are given by%
\begin{align}
\wp^{j,r}\left(  \sigma\right)   & =-\wp^{j-1,r-1}\left(  \sigma\right)
,\text{ \ \ \ \ \ \ }0\leq\sigma\leq\pi/2\\
\wp_{M}^{1}+\wp_{M}^{2}+\wp_{M}^{3}  & =0
\end{align}
where the mid-point momentum is defined in the usual way $\wp_{M}%
\equiv-i\partial/\partial x_{M}=-i\partial/\partial x_{0}=p_{0}$. The comma
coordinates and their canonical momenta obey the usual commutation relations%
\begin{equation}
\left[  \chi^{j,r}\left(  \sigma\right)  ,\wp^{j,s}\left(  \sigma^{\prime
}\right)  \right]  =i\delta^{rs}\delta\left(  \sigma-\sigma^{\prime}\right)
\text{, \ \ }r,s=L,R
\end{equation}
\ In $Z_{3}$\ Fourier space of the comma,the overlap equations for the half
string coordinates read
\begin{align}
Q^{L}\left(  \sigma\right)   & =eQ^{R}\left(  \sigma\right)  ,\text{
\ \ \ \ \ \ }0\leq\sigma\leq\pi/2\label{coordovlqqbar1}\\
\overline{Q}^{L}\left(  \sigma\right)   & =\bar{e}\overline{Q}^{R}\left(
\sigma\right)  ,\text{ \ \ \ \ \ \ }0\leq\sigma\leq\pi
/2\label{coordovlqqbar1bar}\\
Q_{M}  & =\overline{Q}_{M}=0\label{eqnQMQMBARoverlap}\\
Q^{3,L}\left(  \sigma\right)   & =Q^{3,R}\left(  \sigma\right)  ,\text{
\ \ \ \ \ \ }0\leq\sigma\leq\pi/2\label{coordovlqqbar2}\\
Q_{M}^{3}  & =Q_{M}^{3}%
\end{align}
where equation (\ref{eqnQMQMBARoverlap}) is to be understood as an overlap
equation (i.e., its action on the three vertex is zero). Similarly the
canonical momenta of the half string in the $Z_{3}$ Fourier space of the comma
translate into%
\begin{align}
\wp ^{L}\left(  \sigma\right)   & =-e\wp^{R}\left(
\sigma\right)  ,\text{ \ \ \ \ \ \ }0\leq\sigma\leq\pi/2\label{momdovlqqbarmp}%
\\
\overline{\wp }^{L}\left(  \sigma\right)   & =-\bar{e}\overline
{\wp }^{R}\left(  \sigma\right)  ,\text{ \ \ \ \ \ \ }0\leq\sigma
\leq\pi/2\\
\wp^{3,L}\left(  \sigma\right)   & =-\wp^{3,R}\left(
\sigma\right)  ,\text{ \ \ \ \ \ \ }0\leq\sigma\leq\pi/2\label{momdovlqqbar2}%
\\
P_{M}^{3}  & =0\label{momdovl3barmp}%
\end{align}
The overlap conditions on $Q^{r}\left(  \sigma\right)  $ and  $\wp^{r}\left(
\sigma\right)  $ determine the form of the comma three interaction vertex.
Thus in the $Z_{3}$ Fourier space of the comma the overlap equations separate
into two sets. The half string three vertex%
\[
V_{x}^{HS}\left(  b^{1,r\dag},b^{2,r\dag},b^{3,r\dag}\right)
\]
therefore separates into a product of two pieces one depending on $B^{3,r\dag
}$%
\begin{equation}
B^{3,r}=\frac{1}{\sqrt{3}}\left(  b^{1,r}+b^{2,r}+b^{3,r}\right)  ,\text{
\ \ \ }r=L,R
\end{equation}
and the other one depending on $\left(  B^{r\dag},\overline{B}^{r\dag}\right)
$
\begin{align}
B^{r}  & =\frac{1}{\sqrt{3}}\left(  eb^{1,r}+\bar{e}b^{2,r}+b^{3,r}\right)
,\text{ \ \ \ }r=L,R\\
\overline{B}^{r}  & =\frac{1}{\sqrt{3}}\left(  \bar{e}b^{1,r}+eb^{2,r}%
+b^{3,r}\right)  ,\text{ \ \ \ }r=L,R
\end{align}
Notice that in this notation we have $B_{n}^{r\dag}=\overline{B}_{-n}^{r}$ and
$\overline{B}_{n}^{r\dag}=B_{-n}^{r}$ (where the usual convention
$b_{-n}=b_{n}^{\dag}$ applies). Observe that the first of these equations is
identical to the overlap equation for the identity vertex. Hence, the comma
$3$-Vertex takes the form%
\begin{align}
|V_{Q}^{HS}  & >=\int dQ_{M}d\overline{Q}_{M}dQ_{M}^{3}\delta\left(
Q_{M}\right)  \delta\left(  \overline{Q}_{M}\right)  e^{iP_{M}^{3}Q_{M}^{3}}\\
& \times e^{-\frac{1}{2}\left(  B^{3\dag}\left\vert C\right\vert B^{3\dag
}\right)  -\left(  B^{\dag}\left\vert H\right\vert \overline{B^{\dag}}\right)
}\prod_{r=L,R}\left.  |0>^{3,r}|0>^{r}|\overline{0}>^{r}\right. \nonumber
\end{align}
where $C$ and $H$ are infinite dimensional matrices computed in \cite{6} and
the integration over $Q_{M}^{3}$ gives $\delta\left(  P_{M}^{3}\right)  $.
However $P_{M}^{3}=P_{0}^{3}$ and so $\delta\left(  P_{M}^{3}\right)  $ is the
statements of conservation of momentum at the center of mass of the three
strings. Notice that the comma three interaction vertex separates into a
product of two pieces as anticipated. The vacuum of the three strings, i.e.,
$\prod_{j=1}^{3}|0>^{j,L}|0>^{j,R}$, is however invariant under the $Z_{3}%
$-Fourier transformation. Thus we have $\prod_{r=1}^{2}|0>^{3,r}%
|0>^{r}|\overline{0}>^{r}=\prod_{j=1}^{3}|0>^{j,L}|0>^{j,R}$. If we choose to
substitute the explicit values of the matrices, the above expression reduces
to the simple form
\begin{align}
|V_{x}^{HS}  & >=\int\prod_{i=1}^{3}dx_{M}^{i}\delta\left(  x_{M}^{i}%
-x_{M}^{i-1}\right)  \delta\left(  \sum_{j=1}^{3}p_{M}^{j}\right) \nonumber\\
& \times e^{-\sum_{j=1}^{3}\sum_{n=1}^{\infty}b_{n}^{j,L\dag}b_{n}^{j-1,R\dag
}}\left.  |0>_{123}^{L}|0>_{123}^{R}\right. \label{eqnelgV3}%
\end{align}
where $|0>_{123}^{L,R}$denotes the vacuum in the left (right) product of the
Hilbert space of the three strings. Here $b_{n}^{j,L(R)}$ denotes oscillators
in the $L\left(  R\right)  $ $jth$ string Hilbert space. For simplicity the
Lorentz index ($\mu=0,1,...,25$) and the Minkowski metric $\eta_{\mu\nu}$ used
to contract the Lorentz indices, have been suppressed in equation
(\ref{eqnelgV3}). We shall follow this convention throughout this paper.

Though the form of the comma $3$-Vertex given in equation (\ref{eqnelgV3}) is
quite elegant, it is very cumbersome to relate it directly to the $SCSV$
$3$-Vertex due to the fact that connection between the vacuum in the comma
theory and the vacuum in the $SCSV$ is quite involved. One also needs to use
the change of representation formulas \cite{1}\ to recast the quadratic form
in the half string creation operators in terms of the full string
creation-annihilation operators which \ adds more complications to an already
difficult problem. On the other hand the task could be greatly simplified if
we express the comma vertex in the full string basis. This may be achieved
simply by re expressing the comma overlaps in terms of overlaps in the full
string basis. Moreover, the proof of the Ward-like identities will also
simplify a great deal if the comma $3$-Vertex is expressed in the full string
basis. Before we express the half-string $3$-Vertex is expressed in the full
string basis, we need first to solve the comma overlap equations in
(\ref{coordovlqqbar1}), (\ref{coordovlqqbar2}) and (\ref{momdovlqqbarmp}),
(\ref{momdovlqqbar2}) for the Fourier modes of the comma coordinates and
momenta, respectively. The modes in the $Z_{3}$ Fourier space are given by
\begin{align}
Q_{2n-1}^{r}  & =\frac{1}{\pi\sqrt{2}}\int_{-\pi}^{\pi}Q^{r}\left(
\sigma\right)  \cos\left(  2n-1\right)  \sigma d\sigma\text{,}\\
\overline{Q}_{2n-1}^{r}  & =\frac{1}{\pi\sqrt{2}}\int_{-\pi}^{\pi}\overline
{Q}^{r}\left(  \sigma\right)  \cos\left(  2n-1\right)  \sigma d\sigma
\text{,}\\
Q_{2n-1}^{3,r}  & =\frac{1}{\pi\sqrt{2}}\int_{-\pi}^{\pi}Q^{3,r}\left(
\sigma\right)  \cos\left(  2n-1\right)  \sigma d\sigma
\end{align}
where \ $n=1,2,3,...$, and a similar set for the conjugate momenta. The
overlap equations for the coordinates in (\ref{coordovlqqbar1}) and
(\ref{coordovlqqbar1bar}) and the properties imposed in the Fourier expansion
of the comma coordinates%
\begin{align}
Q^{r}\left(  \sigma\right)   & =Q^{r}\left(  -\sigma\right)  \ \text{\ and
\ }Q^{r}\left(  \sigma\right)  =-Q^{r}\left(  \pi-\sigma\right)  \text{,}\\
\overline{Q}^{r}\left(  \sigma\right)   & =\overline{Q}^{r}\left(
-\sigma\right)  \ \text{\ and \ }\overline{Q}^{r}\left(  \sigma\right)
=-\overline{Q}^{r}\left(  \pi-\sigma\right)  \text{,}\\
Q^{3,r}\left(  \sigma\right)   & =Q^{3,r}\left(  -\sigma\right)  \ \text{\ and
\ }Q^{3,r}\left(  \sigma\right)  =-Q^{3,r}\left(  \pi-\sigma\right)  \text{
\ \ }%
\end{align}
where $0\leq\sigma\leq\pi$, imply that their $Z_{3}$ Fourier modes in the
comma basis satisfy%
\begin{align}
Q_{2n-1}^{L}  & =eQ_{2n-1}^{R}\text{,}\label{cmodeovllp2}\\
\overline{Q}_{2n-1}^{L}  & =\bar{e}\overline{Q}_{2n-1}^{R}%
\end{align}
From the overlap in (\ref{coordovlqqbar2}) we obtain
\begin{equation}
Q_{2n-1}^{3,L}=Q_{2n-1}^{3,R}\label{cmodeovllp3}%
\end{equation}
For the Fourier modes of the conjugate momenta one obtains%

\begin{align}
\wp_{2n-1}^{L}  & =-e\wp_{2n-1}^{R},\label{mmodeovllp2}\\
\overline{\wp}_{2n-1}^{L}  & =-\bar{e}\overline{\wp}%
_{2n-1}^{R}%
\end{align}
and%
\begin{equation}
\wp_{2n-1}^{3,L}=-\wp_{2n-1}^{3,R},\text{ \ \ \ \ \ }%
\label{mmodeovllp3}%
\end{equation}
where \ $n=1,2,3,...$. We see that the comma overlaps in the full string basis
separates into a product of two pieces depending on%
\begin{equation}
A_{n}^{3\dag}=\frac{1}{\sqrt{3}}\left(  a_{n}^{1\dag}+a_{n}^{2\dag}%
+a_{n}^{3\dag}\right)
\end{equation}
and on%
\begin{align}
A_{n}^{\dag}  & \equiv A_{n}^{1\dag}=\frac{1}{\sqrt{3}}\left(  \bar{e}%
a_{n}^{1\dag}+ea_{n}^{2\dag}+a_{n}^{3\dag}\right)  ,\\
\overline{A}_{n}^{\dag}  & \equiv A_{n}^{2\dag}=\frac{1}{\sqrt{3}}\left(
ea_{n}^{1\dag}+\bar{e}a_{n}^{2\dag}+a_{n}^{3\dag}\right)  ,
\end{align}
respectively, where the creation and annihilation operators $A_{n}^{\dag}$ and
$A_{n}$ in the $Z_{3}$-Fourier space are defined in the usual way%
\begin{align}
Q_{n}  & =\frac{i}{2}\sqrt{\frac{2}{n}}\left(  A_{n}-A_{n}^{\dag}\right)
,\text{ \ \ }n=1,2,3,...\\
Q_{0}  & =\frac{i}{2}\left(  A_{0}-A_{0}^{\dag}\right) \\
P_{n}  & =-i\frac{\partial}{\partial Q_{n}}=\sqrt{\frac{n}{2}}\left(
A_{n}+A_{n}^{\dag}\right)  ,\text{ \ \ }n=1,2,3,...\\
P_{0}  & =-i\frac{\partial}{\partial Q_{0}}=\left(  A_{0}+A_{0}^{\dag}\right)
\end{align}
and similarly for $\overline{A}_{n}^{\dag}$, $\overline{A}_{n}$ and
$A_{n}^{3\dag}$, $A_{n}^{3}$. Notice that in the $Z_{3}$-Fourier space ,
$A_{n}^{\dag}=\overline{A}_{-n}$, $\overline{A}_{n}^{\dag}=A_{-n}$. For the
matter sector, the comma $3$-Vertex would be represented as exponential of
quadratic form in the creation operators $A_{n}^{3\dag}$, $A_{n}^{\dag}$ and
$\overline{A}_{m}^{\dag}$. Thus the comma $3$-Vertex in the full string
$Z_{3}$-Fourier space takes the form
\begin{equation}
|V_{Q}^{HS}>=\int dQ_{M}d\overline{Q}_{M}\delta\left(  Q_{M}\right)
\delta\left(  \overline{Q}_{M}\right)  V^{HS}\left(  A_{n}^{3\dag},A_{n}%
^{\dag},\overline{A}_{n}^{\dag}\right)  |0)_{123}\label{eqn3-vertexfunctNN}%
\end{equation}
where $|0)_{123}$ denotes the matter part of the vacuum in the Hilbert space
of the three strings and
\begin{equation}
V^{HS}\left(  A_{n}^{3\dag},A_{n}^{\dag},\overline{A}_{n}^{\dag}\right)
=e^{\sum_{n,m=0}^{\infty}\left(  -\frac{1}{2}A_{n}^{3\dag}C_{nm}A_{m}^{3\dag
}-A_{n}^{\dag}F_{nm}\overline{A}_{m}^{\dag}\right)  }%
\label{eqnExp3-vertexfunctNN}%
\end{equation}
The ghost piece of the $3$-Vertex in the bosonized form has the same structure
as the coordinate piece apart from the mid point insertions. In the $Z_{3}%
$-Fourier space $Q_{M}^{\phi}=\overline{Q^{\phi}}_{M}=0$ and only $Q_{M}%
^{\phi3}\neq0$. Thus the mid-point insertion is given by $3iQ_{M}^{\phi3}/2$.
\ The effect of the insertion is to inject the ghost number into the vertex at
its mid-point to conserve the ghost number at the string mid-point, where the
conservation of ghost number is violated due to the concentration of the
curvature at the mid-point. Thus the ghost part of the $3$-Vertex takes the
form%
\begin{equation}
|V_{Q^{\phi}}^{HS}>=e^{3iQ_{M}^{\phi,3}/2}V_{\phi}^{HS}\left(  A_{n}%
^{\phi,3\dag},A_{n}^{\phi\dag},\overline{A^{\phi}}_{n}^{\dag}\right)
|0)_{123}^{\phi}\label{eqnGhostVERT1}%
\end{equation}
where $|0>_{123}^{\phi}$ denotes the ghost part of the vacuum in the Hilbert
space of the three strings and $V_{Q^{\phi}}^{HS}\left(  A_{n}^{\phi,3\dag
},A_{n}^{\phi\dag},\overline{A^{\phi}}_{n}^{\dag}\right)  $ has the exact
structure as the coordinate piece $V_{Q}^{HS}\left(  A_{n}^{3\dag},A_{n}%
^{\dag},\overline{A}_{n}^{\dag}\right)  $. The mid-point insertion
$3iQ_{M}^{\phi,3}/2$ in (\ref{eqnGhostVERT1}) may be written in terms of the
creation annihilation operators%
\begin{equation}
Q_{M}^{\phi,3}=Q_{0}^{\phi,3}+i\sum_{n=even=2}^{\infty}\frac{\left(
-1\right)  ^{n/2}}{\sqrt{n}}\left(  A_{n}^{3}-A_{n}^{3\dag}\right)
\end{equation}
If we now commute the annihilation operators in the mid-point insertion
through the exponential of the quadratic form in the creation operators in the
three-string ghost vertex ($V_{Q^{\phi}}^{HS}$), the three-string ghost vertex
in (\ref{eqnGhostVERT1}) takes the form%
\begin{equation}
|V_{Q^{\phi}}^{HS}>=e^{3iQ_{0}^{\phi,3}/2}e^{3\sum_{n=even=2}^{\infty}%
\frac{\left(  -1\right)  ^{n/2}}{\sqrt{n}}A_{n}^{3\dag}}V_{\phi}^{HS}\left(
A_{n}^{\phi,3\dag},A_{n}^{\phi\dag},\overline{A^{\phi}}_{n}^{\dag}\right)
|0>_{123}^{\phi}%
\end{equation}
We note that commuting the annihilation operators in the mid-point insertion
$3iQ_{M}^{\phi,3}/2$ through $V_{Q^{\phi}}^{HS}\left(  A_{n}^{\phi,3\dag
},A_{n}^{\phi\dag},\overline{A^{\phi}}_{n}^{\dag}\right)  $ results in the
doubling of the creation operator in the mid-point insertion.

\section{The half-string 3-Vertex in the full string basis}

We now proceed to express the half-string overlaps in the Hilbert space of the
full string theory. The change of representation between the half-string modes
and the full string modes derived in \cite{1} is given by%
\begin{align}
Q_{n}^{r}  & =\left(  -1\right)  ^{r+1}Q_{2n-1}+\sum_{m=1}^{\infty}\sqrt
{\frac{2m}{2n-1}}\left[  M_{m\text{ }n\text{ }}^{1}+M_{m\text{ }n\text{ }}%
^{2}\right]  Q_{2m}\nonumber\\
\wp_{n}^{r}  & =\frac{\left(  -1\right)  ^{r+1}}{2}P_{2n-1}+\frac
{1}{2}\sum_{m=1}^{\infty}\sqrt{\frac{2n-1}{2m}}\left[  M_{m\text{ }n\text{ }%
}^{1}-M_{m\text{ }n\text{ }}^{2}\right]  P_{2m}\nonumber\\
& -\frac{\sqrt{2}}{\pi}\frac{\left(  -1\right)  ^{n}}{2n-1}P_{0}%
\end{align}
where $r=1,2\equiv L,R$; $n=1,2,3,...$; and the matrices $M^{1}$ and $M^{2}$
are given by%
\begin{equation}
M_{m\text{ }n\text{ }}^{1}=\frac{2}{\pi}\sqrt{\frac{2m}{2n-1}}\frac{\left(
-1\right)  ^{m+n}}{2m-\left(  2n-1\right)  },\text{ \ \ }%
m,n=1,2,3,...\label{eqnM1}%
\end{equation}
and%
\begin{equation}
M_{m\text{ }n\text{ }}^{2}=\frac{2}{\pi}\sqrt{\frac{2m}{2n-1}}\frac{\left(
-1\right)  ^{m+n}}{2m+\left(  2n-1\right)  },\text{ \ \ }%
m,n=1,2,3,...\label{eqnM2}%
\end{equation}
Now the overlap equations in (\ref{cmodeovllp2}), (\ref{mmodeovllp2}) and
(\ref{eqnQMQMBARoverlap}) become%
\begin{align}
\left(  1+e\right)  Q_{2n-1}  & =-\left(  1-e\right)  \sum_{m=1}^{\infty}%
\sqrt{\frac{2m}{2n-1}}\left[  M_{m\text{ }n\text{ }}^{1}+M_{m\text{ }n\text{
}}^{2}\right]  Q_{2m}\label{cmodeovllp1fmla1-1}\\
\left(  1-e\right)  \frac{1}{2}P_{2n-1}  & =-\left(  1+e\right)  \frac{1}%
{2}\sum_{m=1}^{\infty}\sqrt{\frac{2n-1}{2m}}\left[  M_{m\text{ }n\text{ }}%
^{1}-M_{m\text{ }n\text{ }}^{2}\right]  P_{2m}\nonumber\\
& +\left(  1+e\right)  \frac{\sqrt{2}}{\pi}\frac{\left(  -1\right)  ^{n}%
}{2n-1}P_{0}\label{cmodeovllp1fmla2}\\
Q_{M}  & =Q_{0}+\sqrt{2}\sum_{n=1}^{\infty}\left(  -1\right)  ^{n}%
Q_{2n}=0\label{cmodeovllp1fmla3}%
\end{align}
respectively. The overlaps for the complex conjugate of the first two
equations could be obtained simply by taking the complex conjugation.
Similarly from the overlaps in (\ref{cmodeovllp3}), (\ref{mmodeovllp3}) and
(\ref{momdovl3barmp}) we obtain%
\begin{align}
Q_{2n-1}^{3}  & =0\label{coord3molap}\\
\sum_{m=1}^{\infty}\sqrt{\frac{2n-1}{2m}}\left[  M_{m\text{ }n\text{ }}%
^{1}-M_{m\text{ }n\text{ }}^{2}\right]  P_{2m}^{3}-\frac{2\sqrt{2}}{\pi}%
\frac{\left(  -1\right)  ^{n}}{2n-1}P_{0}^{3}  & =0\label{moord3molap}\\
\wp_{M}^{3}  & =0\label{Midcoord3molap}%
\end{align}
We have seen in reference \cite{1} the  $\wp_{M}^{3}=P_{0}^{3}$ and so the
overlap conditions in (\ref{moord3molap}) and (\ref{Midcoord3molap}) reduce to%
\begin{align}
\sum_{m=1}^{\infty}\sqrt{\frac{2n-1}{2m}}\left[  M_{m\text{ }n\text{ }}%
^{1}-M_{m\text{ }n\text{ }}^{2}\right]  P_{2m}^{3}  & =0\label{eqnoverlappmod}%
\\
P_{0}^{3}  & =0\label{eqnconsofmom}%
\end{align}
It is important to keep in mind that the equality sign appearing in equations
(\ref{cmodeovllp1fmla1-1}) through (\ref{eqnconsofmom}) is an equality between
action of the operators when acting on the comma vertex except for equation
(\ref{eqnconsofmom}) which is the conservation of the momentum carried by the
third string in the $Z_{3}$ Fourier space.

The comma vertex $|V^{HS}\left(  A_{n}^{3\dag},A_{n}^{\dag},\overline{A}%
_{n}^{\dag}\right)  >$ in the full string basis now satisfies the comma
overlaps in (\ref{cmodeovllp1fmla1-1}), (\ref{cmodeovllp1fmla2}),
(\ref{cmodeovllp1fmla3}), (\ref{coord3molap}), (\ref{eqnoverlappmod}) and
(\ref{eqnoverlappmod}). First let us consider the overlaps in
(\ref{cmodeovllp1fmla1-1}), (\ref{cmodeovllp1fmla2}) and
(\ref{cmodeovllp1fmla3}), i.e.,%
\[
\left[  \left(  1+e\right)  Q_{2n-1}+\left(  1-e\right)  \sum_{m=1}^{\infty
}\sqrt{\frac{2m}{2n-1}}\left(  M_{m\text{ }n\text{ }}^{1}+M_{m\text{ }n\text{
}}^{2}\right)  Q_{2m}\text{ \ }\right]  \text{\ }%
\]%
\begin{equation}
\left\vert V^{HS}\left(  A_{n}^{3\dag},A_{n}^{\dag},\overline{A}_{n}^{\dag
}\right)  \right\rangle =0,\label{lASToLQ1-1N00}%
\end{equation}
\ \
\[
\left[  \left(  1-e\right)  \frac{1}{2}P_{2n-1}+\left(  1+e\right)  \frac
{1}{2}\sum_{m=1}^{\infty}\sqrt{\frac{2n-1}{2m}}\left(  M_{m\text{ }n\text{ }%
}^{1}-M_{m\text{ }n\text{ }}^{2}\right)  P_{2m}\right.  -
\]%
\begin{equation}
\left.  \left(  1+e\right)  \frac{\sqrt{2}}{\pi}\frac{\left(  -1\right)  ^{n}%
}{2n-1}P_{0}\right]  \left\vert V^{HS}\left(  A_{n}^{3\dag},A_{n}^{\dag
},\overline{A}_{n}^{\dag}\right)  \right\rangle =0,\label{lASToLP1-1N1}%
\end{equation}%
\begin{equation}
\left[  Q_{0}+\sqrt{2}\sum_{k=1}^{\infty}\left(  -1\right)  ^{k}Q_{2k}\right]
\left\vert V^{HS}\left(  A_{n}^{3\dag},A_{n}^{\dag},\overline{A}_{n}^{\dag
}\right)  \right\rangle =0\label{lASToLQMid1-1NN}%
\end{equation}
(as well as their complex conjugates), where $n=1,2,3,..$. For the remaining
overlaps, i.e., equations in (\ref{coord3molap}) and (\ref{eqnoverlappmod}),
we have

\bigskip%
\begin{equation}
Q_{2n-1}^{3}\left\vert V^{HS}\left(  A_{n}^{3\dag},A_{n}^{\dag},\overline
{A}_{n}^{\dag}\right)  \right\rangle =0,\label{lASToLQ3}%
\end{equation}%
\begin{equation}
\sum_{m=1}^{\infty}\sqrt{\frac{2n-1}{2m}}\left(  M_{m\text{ }n\text{ }}%
^{1}-M_{m\text{ }n\text{ }}^{2}\right)  P_{2m}^{3}\left\vert V^{HS}\left(
A_{n}^{3\dag},A_{n}^{\dag},\overline{A}_{n}^{\dag}\right)  \right\rangle
=0,\label{lASToLP3}%
\end{equation}%
\begin{equation}
P_{0}^{3}\left\vert V^{HS}\left(  A_{n}^{3\dag},A_{n}^{\dag},\overline{A}%
_{n}^{\dag}\right)  \right\rangle =0\label{lASToLP30}%
\end{equation}
where $n=1,2,3,...$. We notice that these overlaps are identical to the
overlap equations for the identity vertex \cite{4,5,7,8}. Thus%
\begin{equation}
C_{nm}=\left(  -1\right)  ^{n}\delta_{nm}\text{, \ \ \ \ \ }n,m=0,1,2,...
\end{equation}
The explicit form of the matrix $F$, may be obtained from the overlap
equations given by (\ref{lASToLQ1-1N00}), (\ref{lASToLP1-1N1}) and
(\ref{lASToLQMid1-1NN}) as well as their complex conjugates. It will turn out
that the matrix $F$ has the following properties%
\begin{equation}
F=F^{\dag},\text{ \ \ \ }\overline{F}=CFC,\text{ \ \ \ }F^{2}%
=1\label{eqnPropertiesofF}%
\end{equation}
which are consistent with the properties of the coupling matrices in Witten's
theory of open bosonic strings \cite{7,8}. This indeed is a nontrivial check
on the validity of the comma approach to the theory of open bosonic strings.

Now substituting (\ref{eqnExp3-vertexfunctNN}) into (\ref{lASToLQ1-1N00}) and
writing $Q_{n}$ in terms of $A_{n}^{\dag}$ and $A_{n}$, we obtain the first
equation for the matrix $F$%

\begin{equation}
F_{2n-1\text{ }k}+\delta_{2n-1\text{ }k}-i\sqrt{3}\sum_{m=1}^{\infty}\left(
M_{m\text{ }n\text{ }}^{1}+M_{m\text{ }n\text{ }}^{2}\right)  \left(
F_{2m\text{ }k}+\delta_{2m\text{ }k}\right)  =0\label{eqnconstr1-1}%
\end{equation}
where $k=0,1,2...,n=1,2,3,...$. Next from the overlap equation in
(\ref{lASToLP1-1N1}) we obtain a second condition on the $F$ matrix%
\begin{align}
0  & =\left(  F_{2n-1\text{ }k}-\delta_{2n-1\text{ }k}\right)  +\frac{1}%
{\sqrt{3}}i\sum_{m=1}^{\infty}\left(  M_{m\text{ }n\text{ }}^{1}-M_{m\text{
}n\text{ }}^{2}\right)  \left(  F_{2m\text{ }k}-\delta_{2m\text{ }k}\right)
\nonumber\\
& -\frac{4}{\pi}\frac{i}{\sqrt{3}}\frac{\left(  -1\right)  ^{n}}{\left(
2n-1\right)  ^{3/2}}\left(  F_{0\text{ }k}-\delta_{0\text{ }k}\right)
\label{eqnconstr1-22}%
\end{align}
where $k=0,1,2...,n=1,2,3,..$. The overlaps for the mid-point in
(\ref{lASToLQMid1-1NN}) give%
\begin{equation}
\left[  \left(  F_{0m}+\delta_{0m}\right)  +\sqrt{2}\sum_{k=1}^{\infty}\left(
-1\right)  ^{k}\sqrt{\frac{2}{2k}}\left(  F_{2km}+\delta_{2k\text{ }m}\right)
\right]  =0,\text{ \ }m=0,1,2,...\label{eqnconstr1-3MP}%
\end{equation}
Solving equations (\ref{eqnconstr1-1}) and (\ref{eqnconstr1-22}), we have%
\begin{equation}
F_{2n\text{ }0}=\frac{1}{\pi}\left(  F_{00}-1\right)  \sum_{m=1}^{\infty
}\left[  \left(  M_{1}^{T}+\frac{1}{2}M_{2}^{T}\right)  ^{-1}\right]
_{nm}\frac{\left(  -\right)  ^{m}}{\left(  2m-1\right)  ^{3/2}}\text{
}\label{F2n0}%
\end{equation}%
\begin{align}
F_{2n\text{ }2k}  & =\frac{1}{\pi}F_{0\text{ }2k\text{ }}\sum_{m=1}^{\infty
}\left[  \left(  M_{1}^{T}+\frac{1}{2}M_{2}^{T}\right)  ^{-1}\right]
_{n\text{ }m}\frac{\left(  -\right)  ^{m}}{\left(  2m-1\right)  ^{3/2}%
}\nonumber\\
& -\sum_{m=1}^{\infty}\left[  \left(  M_{1}^{T}+\frac{1}{2}M_{2}^{T}\right)
^{-1}\right]  _{n\text{ }m}\left[  \frac{1}{2}M_{1}^{T}+M_{2}^{T}\right]
_{m\text{ }k}\text{ }\label{F2n2k}%
\end{align}%
\begin{align}
F_{2n\text{ }2k-1}  & =-\frac{i\sqrt{3}}{2}\left[  \left(  M_{1}^{T}+\frac
{1}{2}M_{2}^{T}\right)  ^{-1}\right]  _{n\text{ }k}+\frac{1}{\pi}F_{0\text{
}2k-1\text{ }}\nonumber\\
& \times\sum_{m=1}^{\infty}\left[  \left(  M_{1}^{T}+\frac{1}{2}M_{2}%
^{T}\right)  ^{-1}\right]  _{n\text{ }m}\frac{\left(  -\right)  ^{m}}{\left(
2m-1\right)  ^{3/2}}\label{F2n2k-1}%
\end{align}%
\begin{align}
F_{2n-1\text{ }2k-1}  & =\frac{2i}{\sqrt{3}}\sum_{m=1}^{\infty}\left[
\frac{1}{2}M_{1}^{T}+M_{2}^{T}\right]  _{n\text{ }m}\text{ }F_{2m\text{ }%
2k-1}\nonumber\\
& +\frac{2i}{\pi\sqrt{3}}\frac{\left(  -\right)  ^{n}}{\left(  2n-1\right)
^{3/2}}F_{0\text{ }2k-1\text{ }}\label{F2n-12k-1}%
\end{align}%
\begin{align}
F_{2n-1\text{ }2k}  & =\frac{2i}{\sqrt{3}}\sum_{m=1}^{\infty}\left[  \frac
{1}{2}M_{1}^{T}+M_{2}^{T}\right]  _{n\text{ }m}\text{ }F_{2m\text{ }2k}%
+\frac{2i}{\sqrt{3}}\left[  M_{1}^{T}+\frac{1}{2}M_{2}^{T}\right]
_{nk}\nonumber\\
& +\frac{2i}{\pi\sqrt{3}}\frac{\left(  -\right)  ^{n}}{\left(  2n-1\right)
^{3/2}}F_{0\text{ }2k\text{ }}\label{F2n-12k}%
\end{align}%
\begin{align}
F_{2n-1\text{ }0}  & =\frac{2i}{\sqrt{3}}\sum_{m=1}^{\infty}\left[  \frac
{1}{2}M_{1}^{T}+M_{2}^{T}\right]  _{n\text{ }m}\text{ }F_{2m\text{ }0}%
+\frac{2i}{\pi\sqrt{3}}\nonumber\\
& \times\frac{\left(  -\right)  ^{n}}{\left(  2n-1\right)  ^{3/2}}\left(
F_{0\text{ }0\text{ }}-1\right) \label{F2n-10}%
\end{align}
where all $n,k=1,2,3,.....$Finally equation (\ref{eqnconstr1-3MP}) leads to%
\begin{align}
\left(  F_{00}+1\right)   & =2\sum_{n=1}^{\infty}\frac{\left(  -1\right)
^{n+1}}{\sqrt{2n}}F_{2n\text{ }0}\text{,}\label{eqnConstF00}\\
F_{0\text{ }2m}  & =2\frac{\left(  -1\right)  ^{m+1}}{\sqrt{2m}}+2\sum
_{k=1}^{\infty}\frac{\left(  -1\right)  ^{k+1}}{\sqrt{2k}}F_{2k\text{ }%
2m}\text{,}\label{eqnQmidF02m}\\
F_{0\text{ }2m-1}  & =2\sum_{k=1}^{\infty}\frac{\left(  -1\right)  ^{k+1}%
}{\sqrt{2k}}F_{2k\text{ }2m-1}\label{eqnQmidF02m-1}%
\end{align}
where $m=1,2,3,...$.

Now the explicit form of the $F$ matrix is completely given by the set of
equations (\ref{F2n0}), (\ref{F2n2k}), (\ref{F2n2k-1}), (\ref{F2n-12k-1}),
(\ref{F2n-12k}), (\ref{F2n-10}), (\ref{eqnConstF00}), (\ref{eqnQmidF02m}) and
(\ref{eqnQmidF02m-1}) provided that the inverse of the $\left(  M_{1}%
^{T}+\frac{1}{2}M_{2}^{T}\right)  $ exist. Now we proceed to compute the
required inverse.

\section{Finding the inverse}

In the half string formulation the combination $\left(  M_{1}^{T}+\frac{1}%
{2}M_{2}^{T}\right)  ^{-1}$ is a special case of the more general expression,
$M_{1}^{T}+\cos\left(  k\pi/N\right)  M_{2}^{T}$, where $k=1,2,3,...,2N$ and
$N$ is the number of strings\footnote{The reason we are considering this more
general expression is that this combination appears in computing the
$N-$interaction vertex which will probe useful in future work and it does not
add to the level of difficulty in finding the inverse.}. For the case of
interest, $N$ corresponds to $3$ and $k=1$. It is however more constructive to
consider the generic combination $\beta M_{1}^{T}+\alpha M_{2}^{T}$. Again for
the case of interest one has $\beta=1$ and $\alpha=\cos\left(  k\pi/N\right)
=1/2$. For the inverse of $\beta M_{1}^{T}+\alpha M_{2}^{T}$, we propose the
Ansatz%
\begin{align}
\left[  \left(  \beta M_{1}^{T}+\alpha M_{2}^{T}\right)  ^{-1}\right]  _{nm}
& =(-)^{n+m}\sqrt{2n}\left[  \alpha^{\prime}\frac{u_{2n}^{1-1/p}u_{2m-1}%
^{1/p}+u_{2n}^{1/p}u_{2m-1}^{1-1/p}}{2n-\left(  2m-1\right)  }\right.
\nonumber\\
& \left.  +\beta^{\prime}\frac{u_{2n}^{1-1/p}u_{2m-1}^{1/p}-u_{2n}%
^{1/p}u_{2m-1}^{1-1/p}}{2n+\left(  2m-1\right)  }\right]  \sqrt{2m-1}%
\label{ansatz}%
\end{align}
The coefficients $u_{k}^{1/p}$ and $u_{k}^{1-1/p}$ are the modes appearing in
the Taylor expansion of the functions $\left(  \frac{1+x}{1-x}\right)  ^{1/p}$
and $\left(  \frac{1+x}{1-x}\right)  ^{1-1/p}$ respectively. For the three
interaction vertex $p=3$ and the Taylor modes $u_{k}^{1/p}$ and $u_{k}%
^{1-1/p}$ reduce to $u_{k}^{1/3}=a_{k}$ and $u_{k}^{2/3}=b_{k}$ found in
references \cite{7,8}. These coefficients are treated in details in appendix
A. The free parameters $\alpha^{\prime}$, $\beta^{\prime}$ and $p$ are to be
determined by demanding that (\ref{ansatz}) satisfies the identities%
\begin{equation}
\left(  \beta M_{1}^{T}+\alpha M_{2}^{T}\right)  ^{-1}\left(  \beta M_{1}%
^{T}+\alpha M_{2}^{T}\right)  =I\label{IdentityeqLI}%
\end{equation}
which implies that $\left(  \beta M_{1}^{T}+\alpha M_{2}^{T}\right)  ^{-1}$ is
left inverse and the identity%
\begin{equation}
\left(  \beta M_{1}^{T}+\alpha M_{2}^{T}\right)  \left(  \beta M_{1}%
^{T}+\alpha M_{2}^{T}\right)  ^{-1}=I\label{IdentityeqRI}%
\end{equation}
which implies that $\left(  \beta M_{1}^{T}+\alpha M_{2}^{T}\right)  ^{-1}$ is
right inverse. Here $I$ is the identity matrix in the space of $N$ strings.

Before we proceed to fix the constants $\alpha^{\prime}$, $\beta^{\prime} $
and $p$, there are two special cases where the inverse could be obtained with
ease with the help of the commutation relations of the half string creation
annihilation operators $\left(  b_{n}^{\left(  r\right)  },b_{n}^{\left(
r\right)  \dag}\right)  $. They are given by $k=2N$ and $k=N$.

For $k=2N$, the combination $M_{1}^{T}+\cos\left(  k\pi/N\right)  M_{2}^{T}$
reduces to $M_{1}^{T}+M_{2}^{T}$ and the inverse $\left(  M_{1}^{T}+M_{2}%
^{T}\right)  ^{-1}=M_{1}-M_{2}$. To see this we only need to verify that
$\left(  M_{1}^{T}+M_{2}^{T}\right)  \left(  M_{1}-M_{2}\right)  =\left(
M_{1}-M_{2}\right)  \left(  M_{1}^{T}+M_{2}^{T}\right)  =I$. We first consider%
\begin{equation}
\left(  M_{1}^{T}+M_{2}^{T}\right)  \left(  M_{1}-M_{2}\right)  =\left(
M_{1}^{T}M_{1}-M_{2}^{T}M_{2}\right)  -\left(  M_{1}^{T}M_{2}-M_{2}^{T}%
M_{1}\right) \label{Proinverse}%
\end{equation}
Using the commutation relations%
\begin{equation}
\left[  b_{n}^{\left(  r\right)  },b_{-m}^{\left(  s\right)  }\right]
=\delta^{rs}\delta_{n+m\text{ }0}\label{Comm2}%
\end{equation}
(where $b_{-m}^{\left(  s\right)  }\equiv b_{m}^{\left(  s\right)  \dag})$ for
the half string creation annihilation modes $\left(  b_{n}^{\left(  r\right)
},b_{n}^{\left(  r\right)  \dag}\right)  ,$ one can show that the the
combination inside the first bracket is the identity matrix \ $I$ and the
combination inside the second bracket is identically zero. To see this recall
that the change of representation between the full string creation
annihilation modes $\left(  a_{n},a_{n}^{\dag}\right)  $ and the half string
creation annihilation modes is given by
\begin{equation}
b_{n}^{\left(  r\right)  }=\left(  -1\right)  ^{r}a_{2n-1}+\frac{1}{2}%
\sum_{m=1}^{\infty}\left(  M_{mn}^{1}a_{2m}-M_{mn}^{2}a_{-2m}\right)  \text{,
\ \ \ \ }n=1,2,3,...\label{FulltoHalf2}%
\end{equation}
and $b_{-n}^{\left(  r\right)  }\equiv b_{n}^{\left(  r\right)  \dag}$ is
given by the same expression with $a_{k}\rightleftharpoons a_{-k}$.
Substituting (\ref{FulltoHalf2})\ into (\ref{Comm2}) one obtains
\begin{align}
0  & =\left[  b_{n}^{\left(  r\right)  },b_{q}^{\left(  s\right)  }\right]
=-\frac{1}{2}\delta^{rs}\sum_{k=1}^{\infty}\left[  \left(  M_{1}^{T}\right)
_{nk}\left(  M_{2}\right)  _{kq}-\left(  M_{2}^{T}\right)  _{nk}\left(
M_{1}\right)  _{kq}\right] \nonumber\\
& =-\frac{1}{2}\delta^{rs}\left[  M_{1}^{T}M_{2}-M_{2}^{T}M_{1}\right]
_{nq}\label{comm3}%
\end{align}
for $n>0$ and $m=-q<0$. Since $\delta^{rs}=0$ for $r\neq s$, then the above
equation does not yield any information about the combination $M_{1}^{T}%
M_{2}-M_{2}^{T}M_{1}$ for\ $r\neq s$. However, for $r=s$, equation
(\ref{comm3}) yields%
\begin{equation}
M_{1}^{T}M_{2}-M_{2}^{T}M_{1}=0\label{Identity1}%
\end{equation}
Similarly one has%
\begin{align}
\delta^{rs}\delta_{nm}  & =\left[  b_{n}^{\left(  r\right)  },b_{-m}^{\left(
s\right)  }\right]  =\frac{1}{2}\left(  -1\right)  ^{r+s}\delta_{nm}%
\nonumber\\
& +\frac{1}{2}\sum_{m=1}^{\infty}\left[  \left(  M_{1}^{T}\right)
_{nk}\left(  M_{1}\right)  _{km}-\left(  M_{2}^{T}\right)  _{nk}\left(
M_{2}\right)  _{km}\right] \nonumber\\
& =\frac{1}{2}\left(  -1\right)  ^{r+s}I_{nm}+\frac{1}{2}\left[  M_{1}%
^{T}M_{1}-M_{2}^{T}M_{2}\right]  _{nm}%
\end{align}
for $n,m>0$. In this case, the above expression gives the following identity%
\begin{equation}
M_{1}^{T}M_{1}-M_{2}^{T}M_{2}=I\label{Identity2}%
\end{equation}
for all possible values of $r$ and $s$. Substituting equations
(\ref{Identity1}) and (\ref{Identity2}) into equation (\ref{Proinverse}) we
arrive at%
\begin{equation}
\left(  M_{1}^{T}+M_{2}^{T}\right)  \left(  M_{1}-M_{2}\right)
=I\label{EsIdentity1}%
\end{equation}
Thus $\left(  M_{1}-M_{2}\right)  $ is the right inverse of $\left(  M_{1}%
^{T}+M_{2}^{T}\right)  $. To complete the proof one must show that $\left(
M_{1}-M_{2}\right)  $ is also a left inverse; that is we need to establish the
following identity%
\begin{equation}
\left(  M_{1}-M_{2}\right)  \left(  M_{1}^{T}+M_{2}^{T}\right)
=I\label{Identity3}%
\end{equation}
The proof of the above identity follows at once from the change of
representation between the half string creation annihilation modes $\left(
b_{n}^{\left(  r\right)  },b_{n}^{\left(  r\right)  \dag}\right)  $ and the
full string creation annihilation modes $\left(  a_{n},a_{n}\dag\right)  $
given by%
\begin{equation}
a_{2n}=\frac{\left(  -1\right)  ^{n}}{\sqrt{2n}}P+\sum_{m=1}^{\infty}\left(
M_{nm}^{1}b_{m}^{\left(  +\right)  }-M_{nm}^{2}b_{-2m}^{\left(  +\right)
}\right)  \text{, \ \ \ \ }n=1,2,3,...\label{HalftoF}%
\end{equation}
(where $b_{n}^{\left(  +\right)  }=\frac{1}{\sqrt{2}}\left(  b_{n}^{\left(
1\right)  }+b_{n}^{\left(  2\right)  }\right)  $) and the commutation
relations%
\begin{equation}
\left[  a_{n},a_{-m}\right]  =\delta_{n+m\text{ }0}.\label{comm4}%
\end{equation}
Using equations (\ref{HalftoF}) and (\ref{comm4}) and skipping the algebraic
details, one obtains the following identities%
\begin{align}
M_{1}M_{2}^{T}-M_{2}M_{1}^{T}  & =0\\
M_{1}M_{1}^{T}-M_{2}M_{2}^{T}  & =I
\end{align}
needed to prove that the combination $\left(  M_{1}-M_{2}\right)  $ is also a
left inverse. This completes the proof.

For $k=N$, the combination $M_{1}^{T}+\cos\left(  k\pi/N\right)  M_{2}^{T}$
reduces to $M_{1}^{T}-M_{2}^{T}$ and the inverse $\left(  M_{1}^{T}-M_{2}%
^{T}\right)  ^{-1}=M_{1}+M_{2}$. The proof that $M_{1}+M_{2}$ is the right
inverse follows at once simply by taking the transpose of the already
established identity in (\ref{EsIdentity1}). To show that the combination
$M_{1}+M_{2}$ is also the left inverse of the combination $M_{1}^{T}-M_{2}%
^{T}$ one only needs to take the transpose of (\ref{Identity3}); thus leading
to the desired result.

Now we proceed to fix the constants in (\ref{ansatz}) for $k\neq N,2N$. From
equations (\ref{eqnM1}) and (\ref{eqnM2}), we have%

\begin{equation}
M_{nm}^{1}=\frac{2}{\pi}\sqrt{\frac{2n}{2m-1}}\frac{\left(  -1\right)  ^{n+m}%
}{2n-\left(  2m-1\right)  }%
\end{equation}
and%
\begin{equation}
M_{nm}^{2}=\frac{2}{\pi}\sqrt{\frac{2n}{2m-1}}\frac{\left(  -1\right)  ^{n+m}%
}{2n+\left(  2m-1\right)  }%
\end{equation}
respectively. First we proceed with the identity in (\ref{IdentityeqLI}). If
we could solve for the free parameters $\alpha^{\prime}$,$\beta^{\prime}$, and
$p$ in terms of the known parameters $\alpha$ and $\beta$ then the Ansatz in
(\ref{ansatz}) is the left inverse of the matrix $\beta M_{1}^{T}+\alpha
M_{2}^{T}$. For the off diagonal elements; that is $q\neq n$, the identity in
(\ref{IdentityeqLI}), yields, after much use of the identities in \cite{9},
\begin{align*}
& \alpha^{\prime}\beta\frac{u_{2n}^{1-1/p}O_{-2n}^{u\left(  1,p\right)
}-u_{2n}^{1-1/p}O_{-2q}^{u\left(  1,p\right)  }+u_{2n}^{1/p}O_{-2n}^{u\left(
p-1,p\right)  }-u_{2n}^{1/p}O_{-2q}^{u\left(  p-1,p\right)  }}{2n-2q}\\
& -\alpha^{\prime}\alpha\frac{u_{2n}^{1-1/p}O_{-2n}^{u\left(  1,p\right)
}-u_{2n}^{1-1/p}O_{2q}^{u\left(  1,p\right)  }+u_{2n}^{1/p}O_{-2n}^{u\left(
p-1,p\right)  }-u_{2n}^{1/p}O_{2q}^{u\left(  p-1,p\right)  }}{2n+2q}\\
& +\beta^{\prime}\beta\frac{u_{2n}^{1-1/p}O_{2n}^{u\left(  1,p\right)
}-u_{2n}^{1-1/p}O_{-2q}^{u\left(  1,p\right)  }-u_{2n}^{1/p}O_{2n}^{u\left(
p-1,p\right)  }+u_{2n}^{1/p}O_{-2q}^{u\left(  p-1,p\right)  }}{2n+2q}\\
& -\beta^{\prime}\alpha\frac{u_{2n}^{1-1/p}O_{2n}^{u\left(  1,p\right)
}-u_{2n}^{1-1/p}O_{2q}^{u\left(  1,p\right)  }-u_{2n}^{1/p}O_{2n}^{u\left(
p-1,p\right)  }+u_{2n}^{1/p}O_{2q}^{u\left(  p-1,p\right)  }}{2n-2q}=0
\end{align*}
where the quantities%
\begin{equation}
O_{\pm n=2k}^{u(q,p)}\equiv\sum_{m=2l+1=1}^{\infty}\frac{u_{m}^{q/p}}{\pm
n+m}\text{ , }n\geq0
\end{equation}
have been considered in \cite{9}. The quantities $O_{-n}^{u\left(  q,p\right)
}$ are related to $O_{n}^{u\left(  q,p\right)  }$ through the identity
$O_{-n}^{u\left(  q,p\right)  }=-\cos\left(  q\pi/p\right)  O_{n}^{u\left(
q,p\right)  }$ \cite{9}. The quantity $O_{n}^{u\left(  q,p\right)  }$ has the
value $\left[  \pi/2\sin\left(  q\pi/p\right)  \right]  u_{2n}^{q/p}$
\cite{9}. In order for the right hand side of the above expression to vanish,
the coefficients of $u_{2n}^{1-1/p}u_{2n}^{1/p}$, $u_{2n}^{1-1/p}u_{2q}^{1/p}$
and $u_{2n}^{1/p}u_{2q}^{1-1/p}$ must vanish separately. The vanishing of the
coefficient of the $u_{2n}^{1-1/p}u_{2n}^{1/p}$ can be established explicitly
by substituting the explicit values for $O_{n}^{u\left(  q,p\right)  }$. The
vanishing of the coefficient of $u_{2n}^{1-1/p}u_{2q}^{1/p}$ term leads to the
following conditions on the free parameters%
\begin{align}
\beta^{\prime}\alpha+\alpha^{\prime}\beta\cos\left(  \frac{1}{p}\pi\right)   &
=0\label{Condone}\\
\alpha^{\prime}\alpha+\beta^{\prime}\beta\cos\left(  \frac{1}{p}\pi\right)   &
=0\label{Condtwo}%
\end{align}
The vanishing of the coefficient of $u_{2n}^{1/p}u_{2q}^{1-1/p}$ does not lead
to new conditions on the free parameters but it provides a consistency
condition. The equivalence between the half-string field theory and Witten's
theory of open bosonic strings will guarantee that this consistency condition
will be met. In fact we have verified this requirement explicitly.

For the diagonal elements ($q=n$), the identity in (\ref{IdentityeqLI}), after
much use of the various identities in \cite{9}, yields%
\begin{align}
1  & =\frac{2(-)^{n+n}}{\pi}\left(  2n\right)  ^{1/2}\left(  2n\right)
^{1/2}\left\{  \alpha^{\prime}\beta\left[  u_{2n}^{1-1/p}\widetilde{O}%
_{-2n}^{u\left(  1,p\right)  }+u_{2n}^{1/p}\widetilde{O}_{-2n}^{u\left(
p-1,p\right)  }\right]  \right. \nonumber\\
& -\alpha^{\prime}\alpha\frac{u_{2n}^{1-1/p}O_{-2n}^{u\left(  1,p\right)
}-u_{2n}^{1-1/p}O_{2n}^{u\left(  1,p\right)  }+u_{2n}^{1/p}O_{-2n}^{u\left(
p-1,p\right)  }-u_{2n}^{1/p}O_{2n}^{u\left(  p-1,p\right)  }}{2\left(
2n\right)  }\nonumber\\
& +\beta^{\prime}\beta\frac{u_{2n}^{1-1/p}O_{2n}^{u\left(  1,p\right)
}-u_{2n}^{1-1/p}O_{-2n}^{u\left(  1,p\right)  }-u_{2n}^{1/p}O_{2n}^{u\left(
p-1,p\right)  }+u_{2n}^{1/p}O_{-2n}^{u\left(  p-1,p\right)  }}{2\left(
2n\right)  }\nonumber\\
& \left.  +\beta^{\prime}\alpha\left[  u_{2n}^{1-1/p}\widetilde{O}%
_{2n}^{u\left(  1,p\right)  }-u_{2n}^{1/p}\widetilde{O}_{2n}^{u\left(
p-1,p\right)  }\right]  \right\}
\end{align}
where%
\begin{equation}
\widetilde{O}_{\pm n=2k}^{u(q,p)}=\sum_{m=2l+1=1}^{\infty}\frac{u_{m}^{q/p}%
}{\left(  \pm n+m\right)  ^{2}}%
\end{equation}
has been considered in \cite{9}. Using the explicit values of $O_{n}^{u\left(
q,p\right)  }$, $O_{-n}^{u\left(  q,p\right)  }$, $\widetilde{O}_{n}^{u\left(
q,p\right)  }$ and $\widetilde{O}_{-n}^{u\left(  q,p\right)  }$ which are
given in \cite{9} and imposing the conditions obtained in (\ref{Condtwo}), the
above expression reduces, after a lengthy exercise, otherwise a straight
forward algebra, to%
\begin{equation}
1=\frac{2}{\pi}\left(  2n\right)  \beta^{\prime}\alpha\left[  u_{2n}%
^{1-1/p}\widetilde{S}_{2n}^{\left(  1,p\right)  }-u_{2n}^{1/p}\widetilde{S}%
_{2n}^{\left(  p-1,p\right)  }\right]
\end{equation}
where the quantities $\widetilde{S}_{\pm n}^{\left(  q,p\right)  }$were
introduced in \cite{9}. The above expression may be reduced further by
expressing $\widetilde{S}_{-n}^{\left(  q,p\right)  }$ in terms of
$\widetilde{S}_{n}^{\left(  q,p\right)  }$ through the relations
\begin{align}
\widetilde{S}_{-2n}^{\left(  1,p\right)  }  & =\widetilde{S}_{2n}^{\left(
1,p\right)  }\cos\frac{\pi}{p}+\left[  1+\cos\left(  \frac{\pi}{p}\right)
\right]  \overline{S}_{0}^{\left(  1,p\right)  }S_{2n}^{\left(  1,p\right)
}\\
\widetilde{S}_{-2n}^{\left(  p-1,p\right)  }  & =\widetilde{S}_{2n}^{\left(
p-1,p\right)  }\cos\frac{\left(  p-1\right)  \pi}{p}+\left[  1+\cos\left(
\frac{\left(  p-1\right)  \pi}{p}\right)  \right] \nonumber\\
& \times\overline{S}_{0}^{\left(  p-1,p\right)  }S_{2n}^{\left(  p-1,p\right)
}%
\end{align}
which have been established in \cite{9}. Hence
\begin{align}
1  & =\frac{2}{\pi}\left(  2n\right)  \alpha^{\prime}\beta\left\{
u_{2n}^{1-1/p}\left[  1+\cos\left(  \frac{\pi}{p}\right)  \right]
\overline{S}_{0}^{\left(  1,p\right)  }S_{2n}^{\left(  1,p\right)  }\right.
\nonumber\\
& \left.  +u_{2n}^{1/p}\left[  1-\cos\left(  \frac{\pi}{p}\right)  \right]
\overline{S}_{0}^{\left(  p-1,p\right)  }S_{2n}^{\left(  p-1,p\right)
}\right\} \label{eqniddelements}%
\end{align}
In arriving at the above expression we used the fact that%
\begin{equation}
\alpha^{\prime}\beta\cos\left(  \frac{\left(  p-1\right)  }{p}\pi\right)
-\beta^{\prime}\alpha=\alpha^{\prime}\beta\cos\left(  \frac{1}{p}\pi\right)
+\beta^{\prime}\alpha=0
\end{equation}
Further simplification of (\ref{eqniddelements}) may be achieved by
substituting the explicit values of $\overline{S}_{0}^{\left(  1,p\right)  }$
and $\overline{S}_{0}^{\left(  p-1,p\right)  }$found in \cite{9}. Thus
equation (\ref{eqniddelements}) reduces to%
\begin{align}
1  & =2n\alpha^{\prime}\beta\left\{  u_{2n}^{1-1/p}\left[  1+\cos\left(
\frac{\pi}{p}\right)  \right]  \tan\left(  \frac{1}{p}\frac{\pi}{2}\right)
S_{2n}^{\left(  1,p\right)  }\right. \nonumber\\
& \left.  +S_{2n}^{\left(  1,p\right)  }\left[  1-\cos\left(  \frac{\pi}%
{p}\right)  \right]  \cot\left(  \frac{1}{p}\frac{\pi}{2}\right)
S_{2n}^{\left(  p-1,p\right)  }\right\}  =2n\nonumber\\
& \times\alpha^{\prime}\beta\sin\left(  \frac{\pi}{p}\right)  \left[
u_{2n}^{1-1/p}S_{2n}^{\left(  1,p\right)  }+S_{2n}^{\left(  1,p\right)
}S_{2n}^{\left(  p-1,p\right)  }\right] \label{eqntuffnott}%
\end{align}
To compute the right-hand side of the above expression we need to evaluate the
expression inside the square bracket. We will show that this expression has
the explicit value $2/2n$. Consider the matrix element defined by%
\begin{equation}
W_{mn}=\frac{u_{m}^{1/p}u_{n}^{1-1/p}+u_{m}^{1-1/p}u_{n}^{1/p}}{m+n}%
\label{eqnIDent2N0}%
\end{equation}
The matrix element $W_{mn}$\ satisfies the following recursion relationship,
which may be verified by direct substitution%
\begin{align}
0  & =\left(  n+1\right)  W_{n+1\text{ \ }m}-\left(  n-1\right)  W_{n-1\text{
\ }m}+\left(  m+1\right)  W_{n\text{ \ }m+1}\nonumber\\
& -\left(  m-1\right)  W_{n\text{ \ }m-1}\label{eqnIDent2NI}%
\end{align}
for $m+n=odd$ $\operatorname{integer}$. Letting $\ n\rightarrow2n-1\geq1$,
$m\rightarrow2m\geq2$ in (\ref{eqnIDent2NI}), we obtain%
\begin{align}
0  & =2nW_{2n\text{ \ }2m}-\left(  2n-2\right)  W_{2n-2\text{ \ }2m}+\left(
2m+1\right)  W_{2n-1\text{ \ }2m+1}\nonumber\\
& -\left(  2m-1\right)  W_{2n-1\text{ \ }2m-1}\label{eqnIDent2NII}%
\end{align}
Summing both sides of (\ref{eqnIDent2NII}) over $m$, we have%
\begin{align}
& 2n\sum_{m=0}^{\infty}W_{2n\text{ \ }2m}-\left(  2n-2\right)  \sum
_{m=0}^{\infty}W_{2n-2\text{ \ }2m}\nonumber\\
& =2nW_{2n\text{ \ }0}-\left(  2n-2\right)  W_{2n-2\text{ \ }0}+W_{2n-1\text{
\ }1}\label{eqnIDent2NVI}%
\end{align}
Substituting the explicit values for $2nW_{2n\text{ \ }0}$, $W_{2n-2\text{
\ }0}$ and $W_{2n-1\text{ \ }1}$\ into (\ref{eqnIDent2NVI}) we obtain%
\begin{align}
& 2n\sum_{m=0}^{\infty}W_{2n\text{ \ }2m}-\left(  2n-2\right)  \sum
_{m=0}^{\infty}W_{2n-2\text{ \ }2m}=u_{2n}^{1/p}+u_{2n}^{1-1/p}\nonumber\\
& -u_{2n-2}^{1/p}-u_{2n-2}^{1-1/p}+\frac{2}{2n}u_{2n-1}^{1/p}-\frac{1}%
{2n}\frac{2}{p}u_{2n-1}^{1/p}+\frac{1}{2n}\frac{2}{p}u_{2n-1}^{1-1/p}%
\label{eqnIDent2NVII}%
\end{align}
Recalling the recursion relations for the Taylor modes established in \cite{9}%
\begin{align}
u_{k+1}^{1/p}  & =\frac{1}{(k+1)}\left[  \frac{2}{p}u_{k}^{1/p}+\left(
k-1\right)  u_{k-1}^{1/p}\right] \label{eqnRecRETmodI}\\
u_{k+1}^{1-1/p}  & =\frac{1}{(k+1)}\left[  \frac{2}{p}\left(  p-1\right)
u_{k}^{1-1/p}+\left(  k-1\right)  u_{k-1}^{1-1/p}\right]
\label{eqnRecRETmodII}%
\end{align}
If we now set $k=2n-1$ in the recursion relations in (\ref{eqnRecRETmodI}) and
(\ref{eqnRecRETmodII}) and then rearrange terms, we have%
\begin{align}
\frac{1}{2n}\frac{2}{p}u_{2n-1}^{1/p}  & =u_{2n}^{1/p}-\frac{1}{2n}\left(
2n-2\right)  u_{2n-2}^{1/p},\label{eqnRecRETmodIA}\\
\frac{1}{2n}\frac{2}{p}u_{2n-1}^{1-1/p}  & =-u_{2n}^{1-1/p}+\frac{2}%
{2n}u_{2n-1}^{1-1/p}+\frac{2n-2}{2n}u_{2n-2}^{1-1/p}\label{eqnRecRETmodIIA}%
\end{align}
Substituting (\ref{eqnRecRETmodIA}) and (\ref{eqnRecRETmodIIA}) in the above
equations into (\ref{eqnIDent2NVII}), we find%
\begin{equation}
2n\sum_{m=0}^{\infty}W_{2n\text{ \ }2m}=\left(  2n-2\right)  \sum
_{m=0}^{\infty}W_{2n-2\text{ \ }2m}\label{eqnFINAID-II}%
\end{equation}
Repeated application of the above identity implies that%
\begin{equation}
2n\sum_{m=0}^{\infty}W_{2n\text{ \ }2m}=2\sum_{m=0}^{\infty}W_{2\text{ \ }%
2m}\label{eqnFINAID-III}%
\end{equation}
Substituting the explicit form of $W_{nm\text{ }}$ into the above identity we
have%
\begin{equation}
\left[  u_{2n}^{1/p}S_{2n}^{\left(  p-1,p\right)  }+u_{2n}^{1-1/p}%
S_{2n}^{\left(  1,p\right)  }\right]  =\frac{2}{2n}\left[  u_{2}^{1/p}%
S_{2}^{\left(  p-1,p\right)  }+u_{2}^{1-1/p}S_{2}^{\left(  1,p\right)
}\right] \label{eqnFINAID-III-F}%
\end{equation}
where%
\begin{align}
S_{n}^{\left(  1,p\right)  }  & \equiv\sum_{n+m=even,m=0}^{\infty}\frac
{u_{m}^{1/p}}{n+m},\text{ }\\
\text{\ }S_{n}^{\left(  p-1,p\right)  }  & \equiv\sum_{n+m=even,m=0}^{\infty
}\frac{u_{m}^{1-1/p}}{n+m}\text{\ }%
\end{align}
To complete the proof, it remains to show that the expression inside the
square bracket on the right hand side of equation (\ref{eqnFINAID-III-F}) is
equal to unity. This we do by explicit computation. Consider%
\begin{equation}
u_{2}^{1/p}S_{2}^{\left(  p-1,p\right)  }+u_{2}^{1-1/p}S_{2}^{\left(
1,p\right)  }\label{eqnsqbrtermN}%
\end{equation}
Using the summation formulas for $S_{n}^{\left(  1,p\right)  }$ and
$S_{n}^{\left(  p-1,p\right)  }$, which are given in \cite{9}, the above
expression reduces to
\begin{equation}
u_{2}^{1/p}S_{2}^{\left(  p-1,p\right)  }+u_{2}^{1-1/p}S_{2}^{\left(
1,p\right)  }=1
\end{equation}
and so equation (\ref{eqnFINAID-III-F}) yields%
\begin{equation}
\left[  u_{2n}^{1/p}S_{2n}^{\left(  p-1,p\right)  }+u_{2n}^{1-1/p}%
S_{2n}^{\left(  1,p\right)  }\right]  =\frac{2}{2n}\label{eqnuSpuSp-1}%
\end{equation}
Substituting this result for the expression in the square bracket in
(\ref{eqntuffnott}) leads to one more condition on the parameters
$\alpha^{\prime}$ and $p$%
\begin{equation}
\alpha^{\prime}=\frac{1}{2\sin\left(  \frac{1}{p}\pi\right)  \beta}%
\end{equation}
Collecting all the conditions on the free parameters, and then solving for the
parameters $\alpha^{\prime}$ and $\beta^{\prime}$, and $p$ in terms of the
known parameters $\alpha$ and $\beta$, we find%
\begin{equation}
\frac{\alpha^{2}}{\beta^{2}}=\cos^{2}\left(  \frac{1}{p}\pi\right)  ,\text{
}\alpha^{\prime}=\frac{1}{2\sin\left(  \frac{1}{p}\pi\right)  \beta},\text{
}\beta^{\prime}=-\frac{\cos\left(  \frac{1}{p}\pi\right)  }{2\sin\left(
\frac{1}{p}\pi\right)  \alpha}\label{eqncob1-2-3}%
\end{equation}
The desired expression for the inverse of $\beta M_{1}^{T}+\alpha M_{2}^{T}$,
is therefore given by substituting the values of $\alpha^{\prime}$ and
$\beta^{\prime}$ given by the above expressions into the Ansatz for $\left(
\beta M_{1}^{T}+\alpha M_{2}^{T}\right)  ^{-1}$ in equation (\ref{ansatz}).
Hence,
\begin{align}
& \left[  \left(  \beta M_{1}^{T}+\alpha M_{2}^{T}\right)  ^{-1}\right]
_{nm}\nonumber\\
& =(-)^{n+m}\frac{\sqrt{2n}\sqrt{2m-1}}{2\sin\left(  \frac{1}{p}\pi\right)
}\left[  \frac{1}{\beta}\frac{u_{2n}^{1-1/p}u_{2m-1}^{1/p}+u_{2n}%
^{1/p}u_{2m-1}^{1-1/p}}{2n-\left(  2m-1\right)  }\right. \nonumber\\
& -\left.  \frac{\cos\left(  \frac{1}{p}\pi\right)  }{\alpha}\frac
{u_{2n}^{1-1/p}u_{2m-1}^{1/p}-u_{2n}^{1/p}u_{2m-1}^{1-1/p}}{2n+\left(
2m-1\right)  }\right]
\end{align}
This shows that the above expression is the left inverse. \ To complete the
proof we need to check that the identity in (\ref{IdentityeqRI}) is also
satisfied and leads to the same conditions as in equations (\ref{eqncob1-2-3}%
). This in fact we did verify. The special cases of $k=N$ and $k=2N$ have been
treated earlier. This completes the construction of the inverse for the
general case of the $\left(  \beta M_{1}^{T}+\alpha M_{2}^{T}\right)  $ matrix.

In the particular case of $\left(  M_{1}^{T}+\cos\frac{k\pi}{N}M_{2}%
^{T}\right)  ^{-1}$, the parameters $\alpha=\cos\left(  k\pi/N\right)  $ and
$\beta=1$ respectively, and the above relations in (\ref{eqncob1-2-3}) become
\begin{equation}
\cos^{2}\left(  \frac{\pi}{p}\right)  =\alpha^{2}=\cos^{2}\left(  \frac{k\pi
}{N}\right)  ,\text{ }\alpha^{\prime}=\frac{1}{2\sin\frac{\pi}{p}},\text{
}\beta^{\prime}=-\frac{\cos\frac{\pi}{p}}{2\sin\frac{\pi}{p}\cos\frac{k\pi}%
{N}}\label{codaf13}%
\end{equation}
For the particular case of $\alpha=\cos\left(  k\pi/N\right)  $ and $\beta=1$,
the relations in (\ref{codaf13}) yield%
\begin{align}
\alpha^{\prime}  & =-\beta^{\prime}=\frac{1}{2\sin\left(  \frac{1}{p}%
\pi\right)  }\\
1/p  & =\left\{
\begin{array}
[c]{c}%
k/N\text{, \ \ \ \ }1\leq k\leq N-1\\
\left(  2N-k\right)  /N\text{, \ \ \ \ \ \ }N+1\leq k\leq2N-1
\end{array}
\right.  \text{ }%
\end{align}
If we choose $1\leq k\leq N-1$, then we have
\begin{align}
& \left[  \left(  M_{1}^{T}+\cos\left(  \frac{k\pi}{N}\right)  M_{2}%
^{T}\right)  ^{-1}\right]  _{nm}=(-)^{n+m}\frac{\sqrt{2n}\sqrt{2m-1}}%
{2\sin\left(  \frac{1}{p}\pi\right)  }\\
& \times\left[  \frac{u_{2n}^{1-1/p}u_{2m-1}^{1/p}+u_{2n}^{1/p}u_{2m-1}%
^{1-1/p}}{2n-\left(  2m-1\right)  }-\frac{u_{2n}^{1-1/p}u_{2m-1}^{1/p}%
-u_{2n}^{1/p}u_{2m-1}^{1-1/p}}{2n+\left(  2m-1\right)  }\right] \nonumber
\end{align}
For the case of interest, that is, the three interaction vertex $N=3$ and
$k=1$ so that $p=3$. \ This implies that the Taylor modes $u_{n}^{1/p}$ and
$u_{n}^{1-1/p}$ in the expansion of $\left(  \frac{1+x}{1-x}\right)  ^{1/p}$
and $\left(  \frac{1+x}{1-x}\right)  ^{1-1/p}$ are $a_{n}$ and $b_{n}$ in the
expansion of $\left(  \frac{1+x}{1-x}\right)  ^{1/3}$ and $\left(  \frac
{1+x}{1-x}\right)  ^{2/3}$ encountered in reference \cite{7,8}. Thus the
inverse of $\left(  M_{1}^{T}+\frac{1}{2}M_{2}^{T}\right)  $ now reads
\begin{align}
& \left[  \left(  M_{1}^{T}+\frac{1}{2}M_{2}^{T}\right)  ^{-1}\right]
_{nm}=\frac{1}{\sqrt{3}}(-)^{n+m}\sqrt{2n}\sqrt{2m-1}\nonumber\\
& \times\left[  \frac{a_{2n}b_{2m-1}+b_{2n}a_{2m-1}}{2n-\left(  2m-1\right)
}+\frac{a_{2n}b_{2m-1}-b_{2n}a_{2m-1}}{2n+\left(  2m-1\right)  }\right]
\label{eqnM1+M2INV}%
\end{align}
This is the required inverse needed to finish the construction of the
half-string three interaction vertex in terms of the full-string basis. The
expression in (\ref{eqnM1+M2INV}) is indeed the right and left inverse of
$M_{1}^{T}+\frac{1}{2}M_{2}^{T}$ as can be checked explicitly. See ref.
\cite{9}.

\section{Computing the explicit values of the matrix elements of the $F$
matrix}

To complete the construction of the comma $3$-Vertex
\begin{equation}
|V^{HS}>=\int dQ_{M}^{3}dQ_{M}d\overline{Q}_{M}\delta\left(  Q_{M}\right)
\delta\left(  \overline{Q}_{M}\right)  e^{iP_{0}^{3}Q_{M}^{3}}V^{HS}\left(
A_{n}^{3\dag},A_{n}^{\dag},\overline{A}_{n}^{\dag}\right)  |0,0,0>
\end{equation}
in the $Z_{3}$-Fourier space of the full string, we need the explicit form of
the $F$ matrix. Here we shall give the steps involved in the computation of
the matrix elements of $F$ and relegate many of the technical details to
appendix A. For the purpose of illustration consider $F_{2n0}$. Substituting
the explicit value of $\left(  M_{1}^{T}+\frac{1}{2}M_{2}^{T}\right)  ^{-1}%
$obtained in (\ref{eqnM1+M2INV}) into equation (\ref{F2n0}) gives%
\begin{align}
F_{2n\text{ }0}  & =\frac{1}{\pi}\left(  F_{00}-1\right)  \sum_{m=1}^{\infty
}\frac{1}{\sqrt{3}}(-)^{n+m}\sqrt{2n}\sqrt{2m-1}\cdot\frac{\left(  -1\right)
^{m}}{\left(  2m-1\right)  ^{3/2}}\nonumber\\
& \times\left[  \frac{a_{2n}b_{2m-1}+b_{2n}a_{2m-1}}{2n-\left(  2m-1\right)
}+\frac{a_{2n}b_{2m-1}-b_{2n}a_{2m-1}}{2n+\left(  2m-1\right)  }\right]
\end{align}
where $n=1,2,3,...$. \ Using partial fractions, the above expression becomes
\begin{align}
F_{2n\text{ }0}  & =\frac{1}{\pi}\left(  F_{00}-1\right)  \frac{1}{\sqrt{3}%
}(-)^{n}\frac{\sqrt{2n}}{2n}[2\left(  a_{2n}\right)  O_{0}^{b}-a_{2n}%
O_{-2n}^{b}\nonumber\\
& -b_{2n}O_{-2n}^{a}-a_{2n}O_{2n}^{b}+b_{2n}O_{2n}^{a}]\label{eqnF2nZero-1}%
\end{align}
where the quantities appearing in the above expression are defined have been
evaluated in \cite{9}. Thus substituting the explicit values of these
quantities into (\ref{eqnF2nZero-1}) and combining terms we find%
\begin{equation}
F_{2n\text{ }0}=\left(  F_{00}-1\right)  \frac{(-)^{n}a_{2n}}{\sqrt{2n}%
}\label{eqnF2noGdlastt}%
\end{equation}

The explicit value of the $F_{00}$ may be computed by substituting
(\ref{eqnF2noGdlastt}) into (\ref{eqnConstF00}). Doing that and rearranging
terms we get
\begin{equation}
\left(  F_{00}+1\right)  =-2\left(  F_{00}-1\right)  \sum_{n=1}^{\infty}\text{
}\frac{a_{2n}}{2n}\label{eqnfixF00}%
\end{equation}
The sum appearing on the right-hand side has the value $\left(  3/2\right)
\ln3-2\ln2$, so we obtain
\begin{equation}
\frac{1+F_{00}}{1-F_{00}}=\ln\frac{3^{3}}{2^{4}}\label{eqnEXPVALF00}%
\end{equation}
which gives the explicit value of $F_{00}$ at once. This result is consistent
with that given in \cite{7,8}. To obtain the explicit value of $F_{0\text{
}2m}$, we first need to evaluate the sum over $k$ in equation
(\ref{eqnQmidF02m}), i.e.,%
\[
\sum_{k=1}^{\infty}\frac{\left(  -1\right)  ^{k+1}}{\sqrt{2k}}F_{2k\text{ }%
2m}\text{ }%
\]
where the explicit expression for $F_{2k\text{ }2m}$ in terms of the change of
representation matrices is given by equation (\ref{F2n2k}). Thus substituting
(\ref{F2n2k}) into the above expression we have%
\begin{align}
& \sum_{k=1}^{\infty}\frac{\left(  -1\right)  ^{k+1}}{\sqrt{2k}}F_{2k\text{
}2m}=\sum_{k=1}^{\infty}\frac{\left(  -1\right)  ^{k+1}}{\sqrt{2k}}\frac
{1}{\pi}F_{0\text{ }2m\text{ }}\sum_{l=1}^{\infty}\frac{\left(  -\right)
^{l}}{\left(  2l-1\right)  ^{3/2}}\nonumber\\
& \times\left[  \left(  M_{1}^{T}+\frac{1}{2}M_{2}^{T}\right)  ^{-1}\right]
_{k\text{ }l}-\sum_{k=1}^{\infty}\frac{\left(  -1\right)  ^{k+1}}{\sqrt{2k}%
}\sum_{l=1}^{\infty}\left[  \left(  M_{1}^{T}+\frac{1}{2}M_{2}^{T}\right)
^{-1}\right]  _{k\text{ }l}\nonumber\\
& \times\left[  \frac{1}{2}M_{1}^{T}+M_{2}^{T}\right]  _{lm}%
\end{align}
If we commute\footnote{Since both the sums over $l$ and $k$ are uniformly
convergent, one may perform the sums in any order. We have carried the sums in
the two different orders and found that the result is the same. However, it is
much easier to perform the sum over $k$ first followed by the sum over $l$
rather than the reverse. Here we shall follow the former.} the sums over $k$
and $l$, we get
\begin{align}
\sum_{k=1}^{\infty}\frac{\left(  -1\right)  ^{k+1}}{\sqrt{2k}}F_{2k\text{
}2m}  & =\frac{1}{\pi}F_{0\text{ }2m\text{ }}\sum_{l=1}^{\infty}\sum
_{k=1}^{\infty}\frac{\left(  -\right)  ^{l}}{\left(  2l-1\right)  ^{3/2}%
}\left(  \cdot\cdot\cdot\right) \nonumber\\
& -\sum_{l=1}^{\infty}\sum_{k=1}^{\infty}\left[  \frac{1}{2}M_{1}^{T}%
+M_{2}^{T}\right]  _{l\text{ }m}\left(  \cdot\cdot\cdot\right)
\label{eqnsumovf2k2m}%
\end{align}
where%
\begin{equation}
\left(  \cdot\cdot\cdot\right)  \equiv\frac{\left(  -1\right)  ^{k+1}}%
{\sqrt{2k}}\left[  \left(  M_{1}^{T}+\frac{1}{2}M_{2}^{T}\right)
^{-1}\right]  _{k\text{ }l}%
\end{equation}
Substituting equation (\ref{eqnM1+M2INV}) for the inverse of the combination
$M_{1}^{T}+\left(  1/2\right)  M_{2}^{T}$ into the above expression and
summing over $k$ from $1$ to $\infty$, we obtain%
\begin{equation}
\sum_{k=1}^{\infty}\frac{\left(  -1\right)  ^{k+1}}{\sqrt{2k}}\left[  \left(
M_{1}^{T}+\frac{1}{2}M_{2}^{T}\right)  ^{-1}\right]  _{k\text{ }l}=-\frac
{2}{\sqrt{3}}\frac{(-)^{l}a_{2l-1}}{\left(  2l-1\right)  ^{1/2}}%
\label{eqnSUMovCOM}%
\end{equation}
In arriving at the above result we made use of the identities%
\begin{equation}
\sum_{k=0}^{\infty}\frac{a_{2k}}{2k-\left(  2l-1\right)  }=-\frac{1}{2}%
\sum_{k=0}^{\infty}\frac{a_{2k}}{2k+\left(  2l-1\right)  }=-\frac{1}{2}%
\frac{1}{\sqrt{3}}\pi a_{2l-1}%
\end{equation}
which were derived in \cite{9}. Similar expressions hold for the sums over
$b_{2k}$; see ref. \cite{9}. Now substituting equation (\ref{eqnSUMovCOM})
into (\ref{eqnsumovf2k2m}) gives
\begin{align}
\sum_{k=1}^{\infty}\frac{\left(  -1\right)  ^{k+1}}{\sqrt{2k}}F_{2k\text{ }%
2m}\text{ }  & =-\frac{1}{2}\ln\frac{3^{3}}{2^{4}}F_{0\text{ }2m\text{ }}%
+\sum_{l=1}^{\infty}\frac{(-)^{l}}{\sqrt{3}}\nonumber\\
& \times\frac{2a_{2l-1}}{\sqrt{2l-1}}\left[  \frac{1}{2}M_{1}^{T}+M_{2}%
^{T}\right]  _{l\text{ }m}%
\end{align}
Using the explicit value of $M_{1}$ and $M_{2}$ and rewriting $\ln\left(
3^{3}/2^{4}\right)  $ in terms of $F_{00}$, the above expression becomes%
\begin{align}
\sum_{k=1}^{\infty}\frac{\left(  -1\right)  ^{k+1}}{\sqrt{2k}}F_{2k\text{ }%
2m}\text{ }  & =-\frac{1}{2}\left(  \frac{1+F_{0\text{ }0\text{ }}%
}{1-F_{0\text{ }0\text{ }}}\right)  F_{0\text{ }2m\text{ }}\nonumber\\
& +\frac{\left(  -1\right)  ^{m}\left(  1-a_{2m}\right)  }{\left(  2m\right)
^{1/2}}%
\end{align}
Substituting this result into (\ref{eqnQmidF02m}), we find%
\begin{equation}
F_{0\text{ }2m}=\left(  F_{00}-1\right)  \frac{\left(  -1\right)  ^{m}a_{2m}%
}{\left(  2m\right)  ^{1/2}}\label{eqnlastFo2m}%
\end{equation}
which has the same form as $F_{2m\text{ }0}$ given in (\ref{eqnF2noGdlastt}).
Thus in this case we see that the property $F_{even\text{ }0}=\left(
F^{\dagger}\right)  _{0\text{ }even}$ holds.

Next we consider the evaluation of{\huge \ }$F_{2n-1\text{ }0}$. If we replace
$M_{1},M_{2}$ and $F_{2m\text{ }0}$ in (\ref{F2n-10}) by their explicit
values, given respectively by equation (\ref{eqnM1}), (\ref{eqnM2}) and
(\ref{eqnF2noGdlastt}), we have
\begin{align}
F_{2n-1\text{ }0}  & =\frac{2i}{\sqrt{3}}\frac{2}{\pi}\left(  F_{00}-1\right)
\frac{\left(  -1\right)  ^{n}}{\sqrt{2n-1}}\left[  \frac{1}{2}\sum
_{m=1}^{\infty}\frac{a_{2m}}{2m-\left(  2n-1\right)  }\right. \nonumber\\
& \left.  +\sum_{m=1}^{\infty}\frac{a_{2m}}{2m+\left(  2n-1\right)  }\right]
+\frac{2i}{\pi\sqrt{3}}\frac{\left(  -\right)  ^{n}}{\left(  2n-1\right)
^{3/2}}\left(  F_{0\text{ }0\text{ }}-1\right)
\end{align}
In order to benefit from the results obtained in \cite{9} to help carry out
the sums we first need to extend the range of the sums to include $m=0$. Hence
adding zero in the form $-a_{0}/\left(  2n-1\right)  +a_{0}/\left(
2n-1\right)  $, the above expression becomes
\begin{align}
F_{2n-1\text{ }0}  & =\frac{2i}{\sqrt{3}}\frac{2}{\pi}\left(  F_{00}-1\right)
\frac{\left(  -1\right)  ^{n}}{\sqrt{2n-1}}\left[  \frac{1}{2}\frac{a_{0}%
}{\left(  2n-1\right)  }+\frac{1}{2}\sum_{m=0}^{\infty}\frac{a_{2m}%
}{2m-\left(  2n-1\right)  }\right. \nonumber\\
& \left.  -\frac{a_{0}}{\left(  2n-1\right)  }+\sum_{m=0}^{\infty}\frac
{a_{2m}}{2m+\left(  2n-1\right)  }+\frac{2i\left(  -\right)  ^{n}\left(
F_{0\text{ }0\text{ }}-1\right)  }{\pi\sqrt{3}\left(  2n-1\right)  ^{3/2}%
}\right] \label{eqnF2nzero}%
\end{align}
The sums in the square brackets have been evaluated in \cite{9}. Thus one
finds%
\begin{equation}
F_{2n-1\text{ }0}=i\left(  F_{00}-1\right)  \frac{\left(  -1\right)
^{n}a_{2n-1}}{\sqrt{2n-1}}\text{,}\label{eqnF2n-10N}%
\end{equation}
where $n=1,2,3,...$. To check if the property $F=F^{\dag}$ continue to hold,
we need to compute explicitly the value of $F_{0\text{ }2n-1}$. It is
important to verify that the matrix $F$ is self adjoint for the consistency of
our formulation. The matrix element $F_{0\text{ }2n-1}$ involves the matrix
element $F_{2n\text{ }2k-1}$ which in turn is expressed in terms of the
combination $\left(  M_{1}^{T}+\frac{1}{2}M_{2}^{T}\right)  ^{-1}$ and the
matrix element $F_{0\text{ }2n-1}$ itself. To carry out the calculation,
unfortunately we first need to compute the explicit value of $F_{2n\text{
}2k-1}$. The matrix element $F_{2n\text{ }2k-1}$ is given by (\ref{F2n2k-1})%
\begin{align}
F_{2n\text{ }2k-1}  & =-\frac{i\sqrt{3}}{2}\left[  \left(  M_{1}^{T}+\frac
{1}{2}M_{2}^{T}\right)  ^{-1}\right]  _{n\text{ }k}+\frac{1}{\pi}F_{0\text{
}2k-1\text{ }}\nonumber\\
& \times\sum_{m=1}^{\infty}\left[  \left(  M_{1}^{T}+\frac{1}{2}M_{2}%
^{T}\right)  ^{-1}\right]  _{n\text{ }m}\frac{\left(  -\right)  ^{m}}{\left(
2m-1\right)  ^{3/2}}%
\end{align}
Substituting the explicit value of $\left(  M_{1}^{T}+\frac{1}{2}M_{2}%
^{T}\right)  ^{-1}$ into the above equation and summing over $m$, we find%
\begin{align}
F_{2n\text{ }2k-1}  & =\frac{(-)^{n+k}\sqrt{2n}\sqrt{2k-1}}{2i}\left[
\frac{a_{2n}b_{2k-1}+b_{2n}a_{2k-1}}{2n-\left(  2k-1\right)  }\right.
\nonumber\\
& \left.  +\frac{a_{2n}b_{2k-1}-b_{2n}a_{2k-1}}{2n+\left(  2k-1\right)
}\right]  +\frac{(-)^{n}a_{2n}}{\sqrt{2n}}F_{0\text{ }2k-1}%
\label{eqnF2n2k-1bela}%
\end{align}
where $n=1,2,3,..$. Combining equation (\ref{eqnF2n2k-1bela}) with equation
(\ref{eqnQmidF02m-1}), leads to%
\begin{align}
F_{0\text{ }2m-1}  & =i(-)^{m}\sqrt{2m-1}\sum_{k=1}^{\infty}\left[
\frac{a_{2k}b_{2m-1}+b_{2k}a_{2m-1}}{2k-\left(  2m-1\right)  }\right.
\nonumber\\
& \left.  +\frac{a_{2k}b_{2m-1}-b_{2k}a_{2m-1}}{2k+\left(  2m-1\right)
}\right]  -2F_{0\text{ }2m-1\text{ }}\frac{1}{2}\left(  \frac{1+F_{00}%
}{1-F_{00}}\right) \label{eqnF02m-1Bef}%
\end{align}
To evaluate the sums appearing in (\ref{eqnF02m-1Bef}) we first need to extend
their range to include $k=0$. Doing so and making use of the result of already
established identities in appendix A, equation (\ref{eqnF02m-1Bef}) reduces
to
\begin{equation}
F_{0\text{ }2m-1}=2i(-)^{m}\frac{a_{2m-1}}{\sqrt{2m-1}}-F_{0\text{ }2m-1\text{
}}\left(  \frac{1+F_{00}}{1-F_{00}}\right)
\end{equation}
Solving the above equation for $F_{0\text{ }2m-1}$, we obtain%
\begin{equation}
F_{0\text{ }2m-1}=-i\left(  F_{00}-1\right)  (-)^{m}\frac{a_{2m-1}}%
{\sqrt{2m-1}}\label{eqnF02n-1Final}%
\end{equation}
which is precisely the adjoint of $F_{2m-1\text{ }0}$; see equation
(\ref{eqnF2n-10N}). Thus we have
\begin{equation}
F_{0\text{ }odd}=\left(  F^{\dag}\right)  _{0\text{ }odd}%
\end{equation}
as expected.

The result obtained in (\ref{eqnF02n-1Final}) may be now used to find the
explicit value of $F_{2n\text{ }2m-1}$. Thus substituting equation
(\ref{eqnF02n-1Final}) back into equation (\ref{eqnF2n2k-1bela}), we find%
\begin{align}
F_{2n\text{ }2m-1}  & =\frac{(-)^{n+m}\sqrt{2n}\sqrt{2m-1}}{2i}\left[
\frac{a_{2n}b_{2m-1}+b_{2n}a_{2m-1}}{2n-\left(  2m-1\right)  }\right.
\label{eqnMatEl-F2n,2m-1}\\
& \left.  +\frac{a_{2n}b_{2m-1}-b_{2n}a_{2m-1}}{2n+\left(  2m-1\right)
}\right]  -i\left(  F_{00}-1\right)  \frac{(-)^{n+m}a_{2n}a_{2m-1}}{\sqrt
{2n}\sqrt{2m-1}}\nonumber
\end{align}
where $n,m=1,2,3,..$.

The computation of the matrix element $F_{2n-1\text{ }2m}$ is indeed quite
cumbersome. The difficulty arises from the fact that the defining equation of
$F_{2n-1\text{ }2m}$, which is given by (\ref{F2n-12k}), involves this summing
over the matrix $F_{2m\text{ }2k}$ which is potentially divergent when the
summing index $m$ takes the $k$ value. The limiting procedures involved in
smoothing out the divergence are quite delicate and require careful
consideration. Thus here we shall only give the final result; the details may
be found in \cite{9},%
\begin{align}
F_{2n-1\text{ }2m}  & =-\frac{(-)^{n+m}\sqrt{2m}\sqrt{2n-1}}{2i}\left[
\frac{a_{2m}b_{2n-1}+b_{2m}a_{2n-1}}{2m-\left(  2n-1\right)  }\right.
\label{eqnMatEl-F2n-1,2m}\\
& \left.  +\frac{a_{2m}b_{2n-1}-b_{2m}a_{2n-1}}{2m+\left(  2n-1\right)
}\right]  +i\left(  F_{00}-1\right)  \frac{(-)^{n+m}a_{2m}a_{2n-1}}{\sqrt
{2m}\sqrt{2n-1}}\nonumber
\end{align}
Comparing equations (\ref{eqnMatEl-F2n,2m-1}) and (\ref{eqnMatEl-F2n-1,2m}),
we see that%
\begin{equation}
F_{even\text{ }odd}=\left(  F^{\dag}\right)  _{even\text{ }odd}%
\end{equation}
as expected.

To complete fixing the comma interaction vertex in the full-string basis we
still need to compute the remaining elements, namely $F_{2n\text{ }2m}$ and
$F_{2n-1\text{ }2m-1}$. The computation of the matrices $F_{2n\text{ }2m}$ and
$F_{2n-1\text{ }2m-1}$ involve two distinct cases. The off diagonal case is
given by $n\neq m$ and the diagonal case is given by $n=m$. Though the off
diagonal elements are not difficult to compute, the diagonal elements are
indeed quite involved and they can be evaluated by setting $n=m$ in the
defining equations for $F_{2n\text{ }2m}$ and $F_{2n-1\text{ }2m-1}$ and then
explicitly performing the sums with the help of the various identities we have
established in \cite{9}. An alternative way of computing the diagonal elements
is to take the limit of $n\rightarrow m$ in the explicit expressions for the
off diagonal elements. We have computed the diagonal elements both ways and
obtained the same result which is a non trivial consistency check on our
formalism. For illustration, here we shall compute the diagonal elements by
the limiting process we spoke of as we shall see shortly. But first let us
compute the off diagonal elements. We first consider $F_{2n\text{ }2m}$. From
equation (\ref{F2n2k}), we have%
\begin{align}
F_{2n\text{ }2k}  & =\frac{1}{\pi}F_{0\text{ }2k\text{ }}\sum_{m=1}^{\infty
}\left[  \left(  M_{1}^{T}+\frac{1}{2}M_{2}^{T}\right)  ^{-1}\right]
_{n\text{ }m}\frac{\left(  -\right)  ^{m}}{\left(  2m-1\right)  ^{3/2}%
}\nonumber\\
& -\sum_{m=1}^{\infty}\left[  \left(  M_{1}^{T}+\frac{1}{2}M_{2}^{T}\right)
^{-1}\right]  _{n\text{ }m}\left[  \frac{1}{2}M_{1}^{T}+M_{2}^{T}\right]
_{m\text{ }k}%
\end{align}
Substituting the explicit value of $\left(  M_{1}^{T}+\frac{1}{2}M_{2}%
^{T}\right)  ^{-1}$ and $\frac{1}{2}M_{1}^{T}+M_{2}^{T}$ into the above
equation, we have%
\begin{align}
F_{2n\text{ }2k}  & =F_{0\text{ }2k\text{ }}\frac{(-)^{n}a_{2n}}{\left(
2n\right)  ^{1/2}}-\frac{(-)^{n+k}\left(  2k\right)  ^{1/2}\left(  2n\right)
^{1/2}}{2}\frac{a_{2n}b_{2k}+b_{2n}a_{2k}}{2n+2k}\nonumber\\
& -\frac{2}{\pi}\frac{(-)^{n+k}}{\sqrt{3}}\sqrt{2k}\sqrt{2n}\left\{  \frac
{1}{2}\sum_{m=1}^{\infty}\frac{a_{2n}b_{2m-1}+b_{2n}a_{2m-1}}{\left[
2n-\left(  2m-1\right)  \right]  \left[  2k-\left(  2m-1\right)  \right]
}\right. \nonumber\\
& \left.  +\sum_{m=1}^{\infty}\frac{a_{2n}b_{2m-1}-b_{2n}a_{2m-1}}{\left[
2n+\left(  2m-1\right)  \right]  \left[  2k+\left(  2m-1\right)  \right]
}\right\} \label{eqnF2n2kNewN}%
\end{align}
The difficulty in evaluating the sums arises from the fact in performing these
sums one usually make use of partial fraction to reduce them to the standard
sums treated in \cite{9}; however partial fraction in this case fails due to a
divergence arising from the particular case when $n=m$. Thus to carry our
program through, we first consider the case for which $n\neq k$. For $n\neq
k$, partial fraction can be used to reduce the sums in the above expression to
the standard results obtained in \cite{9} . Skipping \ some rather straight
forward algebra, we find%
\begin{align}
F_{2n\text{ }2k\text{ }}  & =\frac{F_{0\text{ }2k\text{ }}(-)^{n}a_{2n}}%
{\sqrt{2n}}-\frac{(-)^{n+k}\sqrt{2k}\sqrt{2n}}{2}\nonumber\\
& \times\left[  \frac{a_{2n}b_{2k}+b_{2n}a_{2k}}{2n+2k}+\frac{a_{2n}%
b_{2k}-b_{2n}a_{2k}}{2n-2k}\right]
\end{align}
where $n,k=1,2,3,..$, and $n\neq k$. Substituting the value of $F_{0\text{
}2k\text{ }}$, which is given by equation (\ref{eqnlastFo2m}), we have%
\begin{align}
F_{2n\text{ }2k\text{ }}  & =\left(  F_{00}-1\right)  \frac{(-)^{n+k}%
a_{2n}a_{2k}}{\left(  2n\right)  ^{1/2}\left(  2k\right)  ^{1/2}}%
-\frac{(-)^{n+k}\left(  2k\right)  ^{1/2}\left(  2n\right)  ^{1/2}}%
{2}\nonumber\\
& \times\left[  \frac{a_{2n}b_{2k}+b_{2n}a_{2k}}{2n+2k}+\frac{a_{2n}%
b_{2k}-b_{2n}a_{2k}}{2n-2k}\right]
\end{align}
which is the desired result valid for $n,k=1,2,3,..$,and subject to the
condition $n\neq k$. Note that in this case we have%
\begin{equation}
F_{even\text{ }even}=\left(  F^{\dag}\right)  _{even\text{ }even}%
\end{equation}
as expected. As we pointed earlier the diagonal element $F_{2n\text{ }2k\text{
}}$ may be obtained by taking the limit of $k\rightarrow n$ in equation
(\ref{eqnF2n2kNewN}). Hence%
\begin{align}
F_{2n\text{ }2n\text{ }}  & =F_{0\text{ }2n\text{ }}\frac{(-)^{n}a_{2n}%
}{\left(  2n\right)  ^{1/2}}-\frac{a_{2n}b_{2n}+b_{2n}a_{2n}}{4}-\frac{2}{\pi
}\frac{2n}{\sqrt{3}}\nonumber\\
& \times\left[  \frac{1}{2}a_{2n}\widetilde{S}_{-2n}^{b}+\frac{1}{2}%
b_{2n}\widetilde{S}_{-2n}^{a}+a_{2n}\widetilde{S}_{2n}^{b}-b_{2n}%
\widetilde{S}_{2n}^{a}\right]
\end{align}
This result may be simplified further with the help of the following
identities derived in \cite{9}%
\begin{equation}
\widetilde{S}_{-2n}^{a}=\frac{1}{2}\widetilde{S}_{2n}^{a}+\frac{1}{4}\pi
\sqrt{3}S_{2n}^{a}\text{ , }n>0
\end{equation}
and
\begin{equation}
\widetilde{S}_{-2n}^{b}=-\frac{1}{2}\widetilde{S}_{2n}^{b}+\frac{1}{4}\pi
\sqrt{3}S_{2n}^{b}\text{ , }n>0
\end{equation}
Hence%
\begin{align}
F_{2n\text{ }2n\text{ }}  & =F_{0\text{ }2n\text{ }}\frac{(-)^{n}a_{2n}%
}{\left(  2n\right)  ^{1/2}}-\frac{a_{2n}b_{2n}+b_{2n}a_{2n}}{4}\\
& -\frac{2n}{\pi}\frac{1}{2}\left[  \frac{\pi}{2}\left(  a_{2n}S_{2n}%
^{b}+b_{2n}S_{2n}^{a}\right)  +\sqrt{3}\left(  a_{2n}\widetilde{S}_{2n}%
^{b}-b_{2n}\widetilde{S}_{2n}^{a}\right)  \right] \nonumber
\end{align}
The generalization of the plus combination in the square bracket has been
considered before; its value is given explicitly by setting $p=1/3$ in
equation (\ref{eqnuSpuSp-1})
\begin{equation}
b_{2n}S_{2n}^{a}+a_{2n}S_{2n}^{b}=\frac{2}{2n}%
\end{equation}
Using this identity, we obtain%
\begin{align}
F_{2n\text{ }2n}  & =F_{0\text{ }2n\text{ }}\frac{(-)^{n}a_{2n}}{\sqrt{2n}%
}-\frac{1}{2}b_{2n}a_{2n}-\frac{1}{2}\nonumber\\
& -\frac{2n}{\pi}\frac{\sqrt{3}}{2}\left(  a_{2n}\widetilde{S}_{2n}^{b}%
-b_{2n}\widetilde{S}_{2n}^{a}\right)
\end{align}
Using equation (\ref{eqnlastFo2m}) to eliminate $F_{0\text{ }2n}$, the above
expression becomes%

\begin{equation}
F_{2n\text{ }2n}=\left(  F_{00}-1\right)  \frac{a_{2n}a_{2n}}{2n}-\frac{1}%
{2}b_{2n}a_{2n}-\frac{1}{2}-\frac{2n}{\pi}\frac{\sqrt{3}}{2}\left(
a_{2n}\widetilde{S}_{2n}^{b}-b_{2n}\widetilde{S}_{2n}^{a}\right)
\end{equation}
which satisfies the property%
\begin{equation}
F_{even\text{ }even}=\left(  F^{\dag}\right)  _{even\text{ }even}%
\end{equation}
as expected.

Finally we consider the matrix elements $F_{odd\text{ }odd}$ . From equation
(\ref{F2n-12k-1}), we have
\begin{align}
F_{2n-1\text{ }2k-1}  & =\frac{2i}{\sqrt{3}}\sum_{m=1}^{\infty}\left[
\frac{1}{2}M_{1}^{T}+M_{2}^{T}\right]  _{n\text{ }m}F_{2m\text{ }%
2k-1}\nonumber\\
& +\frac{2i}{\pi\sqrt{3}}\frac{\left(  -\right)  ^{n}}{\left(  2n-1\right)
^{3/2}}F_{0\text{ }2k-1\text{ }}\label{eqn2n-12k-1-174}%
\end{align}
The values of $F_{2m\text{ }2k-1}$ and $F_{0\text{ }2k-1\text{ }}$ are given
by equations (\ref{eqnMatEl-F2n,2m-1}) and (\ref{eqnF02n-1Final})
respectively. Hence, substituting the explicit value of $M_{1}^{T}$ and
$M_{2}^{T}$ in (\ref{eqn2n-12k-1-174}) and skipping some rather
straightforward algebra, we find%

\begin{align}
F_{2n-1\text{ }2k-1}  & =\frac{\left(  -1\right)  ^{n+k}\left(  F_{00}%
-1\right)  }{3}\frac{a_{2k-1}a_{2n-1}}{\left(  2n-1\right)  ^{1/2}\left(
2k-1\right)  ^{1/2}}\nonumber\\
& +\frac{(-)^{k+n}\sqrt{2n-1}\sqrt{2k-1}}{2}\frac{a_{2n-1}b_{2k-1}%
+b_{2n-1}a_{2k-1}}{\left(  2n-1\right)  +\left(  2k-1\right)  }+\frac
{2(-)^{k+n}}{\sqrt{3}}\nonumber\\
& \times\frac{\sqrt{2k-1}\sqrt{2n-1}}{\pi}\sum_{m=0}^{\infty}\left[  \frac
{1}{2}\frac{a_{2m}b_{2k-1}+b_{2m}a_{2k-1}}{\left[  2m-\left(  2n-1\right)
\right]  \left[  2m-\left(  2k-1\right)  \right]  }\right. \nonumber\\
& -\left.  \frac{a_{2m}b_{2k-1}-b_{2m}a_{2k-1}}{\left[  2m+\left(
2n-1\right)  \right]  \left[  2m+\left(  2k-1\right)  \right]  }\right]
\label{eqnF2n-12k-1general}%
\end{align}
Now there are two cases to consider $k\neq n$ and $k=n$. For $k\neq n$,
equation (\ref{eqnF2n-12k-1general}) becomes%

\begin{align}
& F_{2n-1\text{ }2k-1}\nonumber\\
& =\frac{\left(  -1\right)  ^{n+k}\left(  F_{00}-1\right)  }{3}\frac
{a_{2k-1}a_{2n-1}}{\sqrt{2n-1}\sqrt{2k-1}}+\frac{(-)^{k+n}\sqrt{2n-1}%
\sqrt{2k-1}}{2}\nonumber\\
& \times\left[  \frac{a_{2n-1}b_{2k-1}+b_{2n-1}a_{2k-1}}{\left(  2n-1\right)
+\left(  2k-1\right)  }+\frac{a_{2n-1}b_{2k-1}-b_{2n-1}a_{2k-1}}{\left(
2n-1\right)  -\left(  2k-1\right)  }\right] \label{F2n-12k-1final}%
\end{align}
where $n,k=1,2,3$ and we have made use of the results in \cite{9} to evaluate
the various sums. Thus for $n\neq k$, we see that
\begin{equation}
F_{odd\text{ }odd}=\left(  F^{\dag}\right)  _{odd\text{ }odd}\,\text{,
\ \ \ \ for }n\neq k
\end{equation}
For $k=n$, equation (\ref{eqnF2n-12k-1general}) becomes%

\begin{align}
F_{2n-1\text{ }2n-1}  & =\frac{1}{3}\left(  F_{00}-1\right)  \frac
{a_{2n-1}a_{2n-1}}{2n-1}+\frac{1}{2}a_{2n-1}b_{2n-1}\nonumber\\
& +\frac{2}{\sqrt{3}}\frac{\left(  2n-1\right)  }{\pi}\left\{  \frac
{b_{2n-1}\widetilde{E}_{-\left(  2n-1\right)  }^{a}+a_{2n-1}\widetilde{E}%
_{-\left(  2n-1\right)  }^{b}}{2}\right. \nonumber\\
& -\left.  \left[  b_{2n-1}\widetilde{E}_{\left(  2n-1\right)  }^{a}%
-a_{2n-1}\widetilde{E}_{\left(  2n-1\right)  }^{b}\right]  \right\}
\label{eqnF2n2nbefore}%
\end{align}
where we have made use of the results in \cite{9} to evaluate the various sums
appearing in the steps leading to the above result. Using the identities
\begin{align}
\widetilde{E}_{-\left(  2n-1\right)  }^{a}  & =\frac{1}{2}\widetilde{E}%
_{2n-1}^{a}+\frac{1}{4}\pi\sqrt{3}S_{2n-1}^{a}\\
\widetilde{E}_{-\left(  2n-1\right)  }^{b}  & =-\frac{1}{2}\widetilde{E}%
_{2n-1}^{b}+\frac{1}{4}\pi\sqrt{3}S_{2n-1}^{b}%
\end{align}
derived in \cite{9}, the above expression becomes%
\begin{align}
F_{2n-1\text{ }2n-1}  & =\frac{1}{3}\left(  F_{00}-1\right)  \frac
{a_{2n-1}a_{2n-1}}{2n-1}+\frac{1}{2}a_{2n-1}b_{2n-1}+\frac{1}{2}\nonumber\\
& +\frac{\sqrt{3}}{2\pi}\left(  2n-1\right)  \left(  a_{2n-1}\widetilde{E}%
_{\left(  2n-1\right)  }^{b}-b_{2n-1}\widetilde{E}_{\left(  2n-1\right)  }%
^{a}\right) \label{eqnF2n2nfinal}%
\end{align}
which is clearly self adjoint. Thus from equations (\ref{F2n-12k-1final}) and
(\ref{eqnF2n2nfinal}) it follows that%
\begin{equation}
F_{odd\text{ \ }odd}=\left(  F^{\dag}\right)  _{odd\text{ \ }odd}%
\end{equation}
as expected. With this result we, establish that $F=F^{\dag}$ as anticipated.

In the original variables, the comma three-string in (\ref{eqn3-vertexfunctNN}%
) can be written in the form%
\begin{equation}
|V_{x}^{HS}>=\delta\left(
{\displaystyle\sum\limits_{r=1}^{3}}
p_{0}^{r}\right)  V^{HS}\left(  \alpha^{1^{\dag}},\alpha^{2\dag},\alpha
^{3\dag}\right)  |0>_{123}\label{eqncomma3-vertexas}%
\end{equation}
where%
\begin{equation}
V^{HS}\left(  \alpha^{1^{\dag}},\alpha^{2\dag},\alpha^{3\dag}\right)
=e^{-\frac{1}{2}\sum_{r,s=1}^{3}\sum_{n,m=0}^{\infty}a_{-n}^{r}%
\mathcal{F}%
_{nm}^{rs}a_{-m}^{s}}\label{eqncomma3-vertexOrigDeg}%
\end{equation}
The matrix elements $%
\mathcal{F}%
_{nm}^{ij}$ may be obtained by comparing (\ref{eqncomma3-vertexOrigDeg}) to
(\ref{eqn3-vertexfunctNN}) . For example consider the terms involving $a^{1}$
and $a^{2}$ in (\ref{eqn3-vertexfunctNN})%
\begin{align}
& -\frac{1}{2}\left(  \frac{1}{3}a_{-n}^{1}C_{nm}a_{-m}^{2}+\frac{1}{3}%
a_{-n}^{2}C_{nm}a_{-m}^{1}\right) \nonumber\\
& -\left(  \frac{1}{3}\bar{e}^{2}a_{-n}^{1}F_{nm}a_{-m}^{2}+\frac{1}{3}%
e^{2}a_{-n}^{2}F_{nm}a_{-m}^{1}\right) \nonumber\\
& =-\frac{1}{2}a_{-n}^{1}\left[  \frac{1}{3}C-\frac{1}{6}\left(
F+F^{T}\right)  +i\frac{1}{6}\sqrt{3}\left(  F-F^{T}\right)  \right]
_{nm}a_{-m}^{2}\nonumber\\
& -\frac{1}{2}a_{-n}^{2}\left[  \frac{1}{3}C-\frac{1}{6}\left(  F+F^{T}%
\right)  -i\frac{1}{6}\sqrt{3}\left(  F-F^{T}\right)  \right]  _{nm}a_{-m}^{1}%
\end{align}
Comparing this result with the terms $-\frac{1}{2}a_{-n}^{1}%
\mathcal{F}%
_{nm}^{12}a_{-m}^{2}$ and $-\frac{1}{2}a_{-n}^{2}%
\mathcal{F}%
_{nm}^{21}a_{-m}^{1}$, we obtain%
\begin{align}%
\mathcal{F}%
^{12}  & =\frac{1}{6}\left[  \left(  2C-F-\overline{F}\right)  +i\sqrt
{3}\left(  F-\overline{F}\right)  \right] \\%
\mathcal{F}%
^{21}  & =\frac{1}{6}\left[  \left(  2C-F-\overline{F}\right)  -i\sqrt
{3}\left(  F-\overline{F}\right)  \right]
\end{align}
where we have used the fact that $F^{T}=\overline{F}$. Likewise one expresses
the remaining matrix element $%
\mathcal{F}%
_{nm}^{rs}$ in terms of the matrix elements $C_{nm}$ and $F_{nm}$ and their
complex conjugates. All in all we have%
\begin{align}%
\mathcal{F}%
& =\frac{1}{3}\left[  \left(  C+F+\overline{F}\right)  \left(
\begin{array}
[c]{ccc}%
1 & 0 & 0\\
0 & 1 & 0\\
0 & 0 & 1
\end{array}
\right)  +\left(  C-\frac{F+\overline{F}}{2}\right)  \left(
\begin{array}
[c]{ccc}%
0 & 1 & 1\\
1 & 0 & 1\\
1 & 1 & 0
\end{array}
\right)  \right. \nonumber\\
& +\left.  i\frac{\sqrt{3}}{2}\left(  F-\overline{F}\right)  \left(
\begin{array}
[c]{ccc}%
0 & 1 & -1\\
-1 & 0 & 1\\
1 & -1 & 0
\end{array}
\right)  \right] \label{MatrixVijelements}%
\end{align}
which is the same result obtained in ref. \cite{7}. Equation
(\ref{MatrixVijelements}) gives completely the comma interaction three vertex
in the full string basis in the representation with oscillator zero modes.

Sometimes it is useful to express the comma vertex in the momentum
representation. For a single oscillator with momentum $p$ and \ creation
operator $\alpha^{\dag}$, the change of basis is accomplished by%
\begin{equation}
|p)=\exp\left(  -\frac{1}{2}p_{0}\overline{p}_{0}+\overline{p}_{0}\alpha
_{0}^{\dag}+\overline{\alpha}_{0}^{\dag}p_{0}-\alpha^{\dag}\overline{\alpha
}^{\dag}\right)  |0)
\end{equation}
with $|0>$ being the oscillator ground state. Thus using the above identity
and equation (\ref{eqnExp3-vertexfunctNN}) one finds the following
representation for the Vertex in the momentum space%
\begin{align}
& \exp\left[  -\frac{1}{2}\sum_{n,m=0}^{\infty}A_{n}^{3\dag}C_{nm}A_{m}%
^{3\dag}-\sum_{n,m=1}^{\infty}A_{n}^{\dag}F_{nm}^{\prime}\overline{A}%
_{m}^{\dag}\right. \nonumber\\
& -\left.  \sum_{n=0}^{\infty}A_{n}^{\dag}F_{n0}^{\prime}\overline{P}_{0}%
-\sum_{n=0}^{\infty}P_{0}F_{0m}^{\prime}\overline{A}_{m}^{\dag}+\frac{1}%
{2}\overline{P}_{0}F_{00}^{\prime}P_{0}\right] \label{eqn3vertinPrep}%
\end{align}
where the prime matrices $F_{nm}^{\prime}$ are related to the unprimed
matrices $F_{nm}$ by%
\begin{align}
F_{00}^{\prime}  & =\frac{1+F_{00}}{1-F_{00}}\\
F_{0n}^{\prime}  & =\frac{F_{0n}}{1-F_{00}}\text{, \ \ }n=1,2,3,...\\
F_{nm}^{\prime}  & =F_{nm}+\frac{F_{n0}F_{0m}}{1-F_{00}}\text{, \ \ }%
n,m=1,2,3,...
\end{align}
The property $F^{2}=1$ in equation (\ref{eqnPropertiesofF}) implies that in
the momentum representation, the $F^{\prime}$ matrix satisfies%
\begin{equation}
\sum_{k=1}^{\infty}F_{nk}^{\prime}F_{km}^{\prime}=\delta_{nm}\text{,
\ \ }n,m=1,2,3,...\label{eqnProfFPrIII}%
\end{equation}
For $n\neq0$, we have $a_{-n}^{r}\equiv\alpha_{-n}^{r}/\sqrt{n}$, and so
equation (\ref{eqn3vertinPrep}) may be written as
\begin{align}
|V_{x}^{HS}  & >=\int%
{\displaystyle\prod}
p_{0}^{r}\exp\left[  \frac{1}{2}\sum_{r,s=1}^{3}\sum_{n,m=1}^{\infty}%
\alpha_{-n}^{r}G_{nm}^{rs}\alpha_{-m}^{s}+\sum_{r,s=1}^{3}p_{0}^{r}G_{0m}%
^{rs}\alpha_{-m}^{s}\right. \nonumber\\
& +\left.  \frac{1}{2}\sum_{r,s=1}^{3}p^{r}G_{00}^{rs}p_{0}^{s}\right]
|0,p)_{123}\label{eqn3vertinPreporignalvaribles}%
\end{align}
where the matrix $G$ is defined through the relation%
\begin{equation}
G_{nm}^{rs}=-\frac{1}{\sqrt{n+\delta_{n0}}}%
\mathcal{F}%
_{nm}^{\prime rs}\frac{1}{\sqrt{m+\delta_{m0}}}%
\end{equation}

The ghost part of the comma vertex in the full string basis has the same
structure as the coordinate one apart from the mid-point insertions
\begin{equation}
|V_{\phi}^{HS}>=e^{\frac{1}{2}i%
{\displaystyle\sum\limits_{r=1}^{3}}
\phi^{r}\left(  \pi/2\right)  }V_{\phi}^{HS}\left(  \alpha^{\phi,1\dag}%
,\alpha^{\phi,2\dag},\alpha^{\phi,3\dag}\right)  0,N_{ghost}=\frac{3}%
{2}>_{123}^{\phi}\label{eqntheghostV}%
\end{equation}
where the $\alpha^{\prime}$s are the bosonic oscillators defined by the
expansion of the bosonized ghost $\left(  \phi\left(  \sigma\right)  ,p^{\phi
}\left(  \sigma\right)  \right)  $ fields and $V_{\phi}^{HS}\left(
\alpha^{\phi,1\dag},\alpha^{\phi,2\dag},\alpha^{\phi,3\dag}\right)  $ is the
exponential of the quadratic form in the ghost creation operators with the
same structure as the coordinate piece of the vertex.

\section{Conclusion}

We have successfully constructed the comma three interaction vertex of the
open bosonic string in terms of the oscillator representation of the full open
bosonic string. The form of the vertex we have obtained for both the matter
and ghost sectors are those obtained in ref. \cite{7,8,10}. This establishes
the equivalence between Witten's 3-interaction vertex of open bosonic strings
and the half string 3-vertex directly without the need for the coherent state
methods employed in ref. \cite{1,11}.

\end{document}